\documentclass[11pt,a4paper]{article}
\usepackage{jheppub}
\usepackage[utf8]{inputenc} 
\usepackage{multirow}    
\usepackage{xcolor}
\usepackage{graphicx}
\usepackage{amsmath}
\usepackage{amssymb}
\usepackage{braket} 
\usepackage{hyperref}
\usepackage{frontespizio}
\usepackage{comment}  
\usepackage{epsfig}
\usepackage{float}
\usepackage{bbold}
\usepackage[title]{appendix}
\usepackage{multirow} 
\usepackage{tikz}
\usetikzlibrary{decorations.pathmorphing}
\usepackage{varwidth}
\usepackage{tikz}
\usetikzlibrary{backgrounds,patterns,calc}
\usepackage{xparse}  
\usepackage{subfigure}  
\usepackage{bbm}   
\usepackage{mathabx}

\title{Advancements in Functorial Homological Mirror Symmetry}

\author[]{Veronica Pasquarella} 

\affiliation{Shanghai Institute for Mathematics and Interdisciplinary Sciences (SIMIS)}
\affiliation{Block A, International Innovation Plaza, No. 657 Songhu Road, \\ 200433 Yangpu District, Shanghai, China }
 
\emailAdd{veronica-pasquarella@simis.cn}

\abstract{\small Mostly inspired by recent work by Katzarkov, Kontsevich, and Sheshmani, combined with previous work by Aganagic, Ooguri, Saulina and Vafa with regard to BPS black hole microstate counting in terms of topological field theory calculations, we will show how these tools can be applied to concrete setups arising from String Theory, and why the formalism of functorial Homological Mirror Symmetry needs to be further developed. A crucial ingredient will turn out being cobordism techniques for evaluating invariants.}

\makeatletter
\gdef\@fpheader{}
\makeatother
\begin{document} 
\maketitle
%\flushbottom 

\small

\section{Introduction and motivation}  

Since its first formulation, Homological Mirror Symmetry (HMS) has undoubtedly opened a remarkable opportunity for String Theory to address deep questions underlying it with powerful mathematical machinery, whose development has spanned more than three decades \cite{Wit,Kont,SYZ,Kontsevich:2010px,Kontsevich:2009xt,Kontsevich:2008fj,Katzarkov:2008hs,Kontsevich:2006jb,Kontsevich:2006ds,Chen:2005zq,Fronsdal:2005mv,Kontsevich:2000yf,Kontsevich:2000er,Kontsevich:1999eg,Barannikov:1997dwc,Kontsevich:1997vb,Kontsevich:1995qf,Kontsevich:1994qz,Hosono:2020gpj,Lian:2011ni,Hosono:2002uj,Hosono:2002uh,Lian:2001tv,Lian:2002dfk,Lian:2000bb,Lian:2000uh,Lian:1999rp,Lian:1999rq,Lian:1998ap,Lian:1997fkq,Seidel:2021drs,Seidel:2019zgf,Seidel:2014rpa,Seidel:2012jj,Seidel:2009qpa,Seidel:2003kb,Abouzaid:2018ika,Abouzaid:2017clg,Abouzaid:2016lym,Abouzaid:2015kaa,Abouzaid:2013ppe,Bocklandt:2011wk,Abouzaid:2010fk,Abouzaid:2004ib,Auroux:2024ika,Cavenaghi:2024xfu,Horja:2022ful,Haiden:2021sqy,Katzarkov:2021glg,Borisov:2019rcs,Kasprzyk:2019ayb,Ballard:2018xyz,Katzarkov:2018jee,Ballard:2016ncw,Katzarkov:2013lsa,Dimitrov:2013xpa,Rebhan:2013rea,Garavuso:2011nz,Kapustin:2009dvd,Chang:2022qmz,Li:2019ifj,Chang:2016opu,Banerjee:2023zqc,Lee:2023piu,Guo:2021aqj,Jockers:2012dk}\footnote{Some additional related treatments are \cite{KKP,DO,PTVV,Brav:2012qsa,Brav:2013fnm,Ben-Bassat:2013nht,Brav:2018byf,YQ,YQ1,Aganagic:2024sww,Gukov:2022gei,Losev:2000bf,Losev:2022tzr,Losev:2023bhj,Losev:2023uxa,Frenkel:2005ku}.}. 

However outstanding the achievements in the field have been, there are still many open questions, both on the Theoretical Physics and Mathematics side of this remarkable correspondence, that necessitate further investigation.  

The present work is part of a much more ambitious project lying at the interface of the two fields, HMS and String Theory, namely identifying the best Calabi-Yau (CY) candidate to embed the Standard Model of Particle Physics in a Unifying Theory of Fundamental Interactions, \cite{Donagi:2000zf,Donagi:2000zs,Donagi:2000vs,Donagi:1999ez,Cicoli:2021dhg}. The present work relies on the trust that such question can be addressed by promoting the mathematical formulation of HMS, specifically in its (generalised) functorial formulation, as a possible selection criterion. The first stages of this project are explained in the present work. However, as we shall see, generalising functorial HMS also constitutes a promising setting for pure mathematics advancements in their own right.

Derived and noncommutative geometry is a paradigm in which to formulate and study various dualities pioneered by String Theory, including Mirror Symmetry. In the context of topological string theory, the relevant physical theory is an extended 2D toplogical field theory (TFT).  A central problem in noncommutative geometry is that of extracting the partition function of the TFT on various manifolds from the data of the category, and to describe various geometric operations on the tangent spaces for the A- and B-model TFT in purely categorical terms. Relying on some recent advancements in this topic, connecting them to calculations in perturbative topological String Theory, we highlight possible directions for further extending the functorial formulation of Homological Mirror Symmetry.
 
As we shall see, two main pillars of our analysis rely on the following correspondences, that have been put forward in the past few decades\footnote{RW and AG in these diagrams are a shorthand notation for Rozansky-Witten and Algebraic Geometry, respectively.}:   

\begin{equation}  
\text{Gauging}  \ \ \ \ \ \ \ \longleftrightarrow\ \ \ \ \ \ \ \text{deformation quantisation}    
\label{eq:defquant}
\end{equation}

\begin{equation}  
\text{RW-model}\ \ \ \ \ \ \ \longleftrightarrow\ \ \ \ \ \ \ \text{Categorified AG}  
\label{eq:catdt}
\end{equation}
How such correspondences fit within the purpose of our work will be explained in due course. Our main finding consists in having related stability and transversality in the degeneracy formula for non-ideal sheaves to nonabelian gauging in the RW-theory setup.

The present work is structured as follows: Section \ref{sec:1} introduces symmetric obstruction theories as shifted symplectic structures, \cite{DO, PTVV}. Section \ref{sec:2} briefly overviews a recent work by Katzarkov, Kontsevich and Sheshmani, \cite{KKS}, pertaining the categorification of Donaldson-Thomas (DT) invariants for cases involving non-smooth moduli spaces, relative CY structures and spherical functors. The example we outline falls within the right-hand-side of \eqref{eq:catdt}. We highlight the types of mathematical structures we need to apply to achieve the degeneracy formula, and how it relates to deformation quantisation. In Section \ref{sec:3} we then turn to the topological string theory formulation of Section \ref{sec:2}, mostly related to \cite{Pan,AOSV}. Here we stress the role played by different brane types in the braneweb for which the topological partition function is being calculated, and how they are encoded in a 2-functor corresponding to Gromov-Witten (GW) invariants\footnote{DT-invariants are an equivariant formulation of GW-invariants. Calculations coincide for CY3-folds.}. Cobordism calculations, though, highlight that such invariants restrict to critical points of a topological Morse function. More interesting configurations should be expected to arise when attempting to analyse geometric structures arising at arbitrary points of the flow. Section \ref{sec:4} applies the tools outlined in the previous sections to a specific setup of interest for its applicability to String Theory, namely that of Moore-Tachikawa varieties, \cite{MT,MS,P,CM}. Particular emphasis is placed on the role of Coulomb branches for classifying 2D Topological Field Theories (TFTs), \cite{XY,CT,CT1,BFN}. The reason for doing so is that such 2D TFTs constitute the boundary of a 3D TFT described by Rozansky and Witten for describing the category of branes associated to the B-model and their deformation. Section \ref{sec:5} turns to a major technique used for extending HMS for singular varieties, namely that of abelianisation. In so doing, we explain the relation between gauging and deformation quantisation, \eqref{eq:defquant}, and the difference between the abelian and the nonabelian setup, highlighting how this relates to the definition of the virtual fundamental class in the degeneracy formula. We conclude by outlining some main open questions and currently ongoing work by the same author towards extending the formalism in question.

\section{Symmetric Obstruction Theories}   \label{sec:1}

The present section is mostly an overview of symmetric obstruction theories and their relation to shifted symplectic structures\footnote{Most of the material outlined in this section is taken from the pioneering works of \cite{DO, PTVV}.}.  This is to set some preliminary tools and notations that will be referred to in later sections.

Symmetric obstruction theories (SOTs), \cite{DO}, are particularly versatile for studying HMS, as we will be explaining in due course. For the time being, we split this first section into two parts: 

\begin{enumerate} 

\item   Section \ref{sec:1.1} outlines some key examples of SOTs as (-1)-shifted symplectic structures.

\item   Section \ref{sec:1.2} explains how the formalism of Section \ref{sec:1.1} can be lifted to higher structures, \cite{PTVV}.

\end{enumerate}

Indeed, examples of SOTs are the following:  

\begin{itemize}  

\item The critical point of a regular function on a smooth variety.   

\item The intersection of Lagrangian submanifolds of a complex symeplectic manifold.  

\end{itemize}    

Both examples are clearly pertinent to HMS applications\footnote{We refer the interested reader to the vast literature in this fascinating subject, pioneered by \cite{Wit,Kont,SYZ}. For completeness, we succinctly summarise some key tools defining the dictionary in appendix \ref{sec:A}.}. Importantly, they can be used to calculate 0-dimensional DT-invariants of projective CY 3-folds, as also addressed by \cite{KKS}. We will be coming back to this in Section \ref{sec:2}, with SOTs being closely tied to Higgs bundle stability\footnote{A brief explanation of which is provided in appendix \ref{sec:B}.}.

\subsection{Shifted symplectic and Lagrangian structures}   \label{sec:1.1}

Three main examples of (-1)-shifted symplectic structures, and thus of symmetric obstruction theories are as follows: 

\begin{itemize}    

\item Sheaves on CY 3-folds.  

\item Maps from elliptic curves to a symplectic target. 

\item Lagrangian intersections.  

\end{itemize}

\subsection*{Obstruction Theory}

 Every obstruction theory practically arises in the following way. For any derived stack, $F$, which is locally of finite presentation, its truncation, $h^{^0}(F)$ comes equipped with a natural perfect obstruction theory. It is constructed by considering the inclusion  

\begin{equation}  
j: \ h^{^0}(F)\ \rightarrow\ F  
\end{equation}   
and by noticing that the induced morphism of cotangent complexes 

\begin{equation}  
j^*: j^*\left(\mathbb{L}_{_{F/k}}\right)\ \rightarrow\ \mathbb{L}_{_{h^{^0}(F)/k}} 
\end{equation}    
satisfies the property of it being a perfect obstruction theory.  

Assuming that $F$ comes equipped with (-1)-shifted symplectic structure $\omega$, we can write the underlying 2-form of degree -1 as a morphism of perfect complexes

\begin{equation}  
\omega:\ \mathbb{T}_{_F}\ \wedge\ \mathbb{T}_{_F}\ \rightarrow\ {\cal O}_{_F}[-1]. 
\end{equation}

Using the fact that $\omega$ is non-degenerate, and that  

\begin{equation}  
\Theta_{_{\omega}}:\ \mathbb{T}_{_F}[-1]  \ \simeq\ \mathbb{L}_{_F}[-1]
\end{equation}  
we get the pairing morphism      

\begin{equation}  
\text{Sym}^{^2}\left(\mathbb{L}_{_F}\right)[-2]\ \simeq\ \left(\mathbb{L}_{_F}[-1]\right)\ \wedge\ \left(\mathbb{L}_{_F}[-1]\right)   \ \simeq\ \mathbb{T}_{_F}\ \wedge\ \mathbb{T}_{_F}\ \rightarrow\ {\cal O}_{_F}[-1]    
\end{equation}  
that can be rewritten as follows  

\begin{equation}  
\text{Sym}^{^2}\left(\mathbb{L}_{_F}\right)\ \rightarrow\ {\cal O}_{_F}[1],   
\end{equation}    
also defined as follows

\begin{equation}  
\mathbb{L}_{_F}\ \simeq\ \mathbb{T}_{_F}[1].     
\end{equation}  

Restricting to the truncation   

\begin{equation}   
h^{^0}\ \hookrightarrow\ F  
\end{equation}   
they find that the perfect obstruction theory  

\begin{equation}  
E\ \overset{def.}{\equiv}\ j^{^*}\left(\mathbb{L}_{_{F/k}}\right)   
\end{equation}    
comes equipped with a natural symmetric equivalence   

\begin{equation}  
E\ \simeq\ E^{^{\text{V}}}[1],    
\end{equation}   
which, by definition is a symmetric obstruction theory.  

For the purpose of the present work, it is particularly instructive to look at two of the examples mentioned earlier on, namely that of sheaves on CY 3-folds, and Lagrangian intersections. We briefly overview them for copleteness, while referring the interested reader to the original work of \cite{PTVV} for a detailed treatment.

\subsubsection*{Sheaves on CY 3-folds}

The first case of interest is that of sheaves on CY 3-folds. Let $X$ be a smooth and proper CY manifold of dimension 3, together with a trivialisation 

\begin{equation}    
\omega_{_{X/k}}\ \simeq\ {\cal O}_{_X}.     
\end{equation}

It is an ${\cal O}$-compact object endowed with an ${\cal O}$-orientation of dimension 3. Then, the derived stack of perfect complexes on $X$, $\mathbb{R}$Perf$(X)$, is classically endowed with a (-1)-shifted symplectic structure, thereby defining a symmetric obstruction theory on the truncation    

\begin{equation}  
h^{^0}\left(\mathbb{R}\text{Perf}(X)\right).   
\end{equation}  

Using the determinant map  

\begin{equation}  
\text{det}:\ \mathbb{R}\text{Perf}(X)\ \rightarrow\ \mathbb{R}\text{Perf}(X),   
\end{equation}
and consider the fiber at a given global point $L\in\text{Pic}(X)$, corresponding to a line bundle on $X$  

\begin{equation}  
\mathbb{R}\text{Perf}(X)_{_L}\ \equiv\ \text{det}^{-1}\left(\{L\}\right).   
\end{equation}   

The (-1)-shifted symplectic form on $\mathbb{R}\text{Perf}(X)$ can be pulled-back to a closed 2-form on $\mathbb{R}\text{Perf}(X)_{_L}$ by means of the natural morphism  

\begin{equation}   
\mathbb{R}\text{Perf}(X)_{_L}\ \rightarrow\ \mathbb{R}\text{Perf}(X).      
\end{equation}     

This closed (-2)-form, stays non-degenerate, implying it defines a (-1)-shifted symplectic structure on $\mathbb{R}\text{Perf}(X)_{_L}$.

\subsubsection*{Lagrangian intersections}    

The other example of interest to us is that if Lagrangian intersections\footnote{We will be coming back to this in Section \ref{sec:5}, in the context of its applicability to extensions of HMS for Coulomb branches of 3D ${\cal N}=4$ SCFTs.}. Given a smooth symplectic scheme $(X,\omega)$, over $\mathbb{k}$, with two smooth Lagrangian schemes $L, L^{\prime}$. Then, the two closed immersions $L, L^{\prime}\subset X$ are endowed with a unique Lagrangian structure, implying 

\begin{equation}  
L\ \underset{X}{\times}\ L^{^{\prime}}  
\label{eq:fiberproduct}
\end{equation}   
carries a natural (-1)-shifted symplectic structure. This also defines a symmetric perfect obstruction theory of amplitude [-1,0] on the truncation, namely on the schematic fiber product \eqref{eq:fiberproduct}.

\subsubsection{PTVV}   \label{sec:1.2}

The second part of this section outlines some main results of the original work by Pantev, T$\ddot{\text{o}}$en, Vaquie, Vezzosi (PTVV), \cite{PTVV}. The main point of their pioneering work consisted in replacing the holomorphic symplectic form by an $n$-shifted symplectic structure on a derived structure. 
Specifically, there are two main theorems that are of particular importance for the present work.

\medskip    

\medskip

\underline{Theorem 1} [PTVV, 2013]  
Let $Y$ be a derived Artin stack, and $X$ an ${\cal O}$-compact derived stack with a $d$-orientation, $[X]$. Assuming the derived mapping stack 

\begin{equation}  
\underline{\text{Map}}(X,Y)  
\end{equation}  
is a local derived Artin stack locally of finite presentation over $\mathbb{k}$, then, we have a transgression map  

\begin{equation}  
\int_{_{[X]}}\text{ev}^{^*}(-):\ \text{Symp}(Y,n)\ \longrightarrow\ \ \text{Symp}\left(\underline{\text{Map}}(X,Y), n-d\right).  
\end{equation}

\medskip    

\medskip 

\underline{Theorem 2} [Calaque, 2015]   

Given $\underline{\text{Map}}(X,Y)$, $\underline{\text{Map}}(X^{^{\prime}},Y)$ derived Artin stacks locally of finite presentation over $\mathbb{k}$, and $i: X\rightarrow X^{^{\prime}}$ a morphism equipped with a nondegenerate boundary structure, then the pullback map  

\begin{equation}  
i^{^*}:\ \underline{\text{Map}}(X,Y^{\prime})\ \longrightarrow\ \underline{\text{Map}}(X,Y) 
\end{equation}
is equipped with a natural Lagrangian structure.

The derived critical locus of any function carries a shifted symplectic structure, since Crit$(f)$ is the intersection of the zero section with the graph of $df$ in the cotangent bundle. 

\medskip    

\medskip 

\underline{Theorem 3} [PTVV, 2013]  
Let $(X,\omega)$ be a derived stack with an $n$-shifted symplectic structure, and let $L\rightarrow X$ and $L^{^{\prime}}\rightarrow X$ be morphisms of derived stacks equipped with Lagrangian structures. Then, $L\underset{X}{\times}L^{^{\prime}}$ is equipped with a natural $(n-1)$-shifted symplectic structure.

The work of PTVV, \cite{PTVV}, certainly opened interesting persepctives for developing the A-side of the HMS correspondence. For example, their technique has been adopted in \cite{CM} to recover and extend the Moore-Tachikawa setup, \cite{MT}. Furthermore, Safronov, \cite{Sa}, and collaborators, have made significant contributions in the context of geometric invariant theory (GIT).

We now breifly turn to explaining how shifted symeplectic structures arise in the GIT setup. Let $M$ be equipped with a 0-shifted symplectic structure and with a Hamiltonian G-action with moment map $\mu$. The stack  

\begin{equation}  
\mu^{^{-1}}(0)/G, 
\end{equation}  
resulting by Hamiltonian reduction of $M$ comes equipped with a canonical 0-shifted symplectic structure. On the other hand, let 

\begin{equation}  
\Lambda_{_1}=0/G\ \rightarrow\ \mathfrak{g}^{*}/G  
\end{equation}  
be the map induced by the embedding   

\begin{equation}  
\{0\}\ \hookrightarrow\ \mathfrak{g}^{*},   
\end{equation}  
and

\begin{equation}  
\Lambda_{_2}=M/G\ \rightarrow\ \mathfrak{g}^{*}/G  
\end{equation} 
be the map induced by $\mu$. One then has a natural isomorphism  

\begin{equation}  
\Lambda_{_1}\ \times_{_{\mathfrak{g}^{*}/G }}\ \Lambda_{_2}\ =\ 0/G\ \times_{_{\mathfrak{g}^{*}/G }}\ M/G\ \simeq\ \mu^{^{-1}}(0)/G.   
\end{equation}   

The stack $\mathfrak{g}^{*}/G$ has the canonical 1-shifted symplectic structure, and each of the two maps  

\begin{equation}  
\Lambda_{_i}\ \longrightarrow\ \mathfrak{g}^{*}/G\ \ \ , \ \ \ i=1,2  
\end{equation}  
has a Lagrangian structure.  The stack 

\begin{equation}  
0/G\ \times_{_{\mathfrak{g}^{*}/G}} M/G  
\end{equation}  
comes equipped with a 0-shifted symplectic structure. More generally, for any stack, $\cal Y$, equipped with an n-shifted symplectic structure and a pair   

\begin{equation}  
\Lambda_{_i}\ \longrightarrow\ {\cal Y}\ \ \ , \ \ \ i=1,2  
\end{equation} 
of Lagrangian structures, the stack  

\begin{equation}  
\Lambda_{_1}\times_{_{\cal Y}}\ \Lambda_{_2}  
\end{equation}  
has a natural (n-1)-shifted symplectic structure.

\section{Categorification of DT-invariants}   \label{sec:2}

We now turn to one of the main applications of SOTs, namely that of categorification of 0D DT-invariants of CY threefolds.

The present section is organised as follows: 

\begin{enumerate}  

\item  At first, we briefly overview the setup of \cite{KKS}, particularly focusing on the need ofr extracting the obstruction theory for singular varieties, stability of the Higgs bundles, and the calculation of DT-invariants.

\item   Section \ref{sec:3.1} presents Tyurin degenerations as an analogue to gluing of oriented manifolds, thereby paving the wasy towards the 2D TFT description of the same setup in subsequent sections of the present work.

\item  Section \ref{sec:3.2} introduces the notion of spherical functors in relation to relative CY structures highlighting the flexibility provided by the noncommutative framework.

\end{enumerate}

We start from briefly overviewing the geometric setup of \cite{KKS}\footnote{Addressing the question of S-duality arising in Superstring Theory.}. Given a smooth projective CY3-variety, $X$. Assuming Pic$(X)$ is generated by an ample divisor $L$. For a fixed integer $k\ \in\ \mathbb{Z}_{_{>0}}$, $H\ \in\ |kL|$, we get a sheaf of $H$-modules over the $H$-hypersurface\footnote{Where D-branes are placed.}, whose fiber jumps in dimension upon crossing $H$.   

We need to define the moduli space of semistable sheaves   

\begin{equation}   
{\cal M}\left(X, \text{Ch}^{^{\bullet}}\right),    
\label{eq:mch}
\end{equation}  
where $\text{Ch}^{^{\bullet}}$ denotes the \emph{Chow homology}. The calculation of the moduli space \eqref{eq:mch} is usually quite involved. Let $\ell$ be the generator of 

\begin{equation}    
H^{^4}(X;\mathbb{Z})\ \simeq\ \mathbb{Z}.  
\end{equation}   

By Poincare' duality, we can define $H^{^2}$.
Given a pair of fixed integers  $i,n\in\mathbb{Z}$, the Chow homology reads as follows  

\begin{equation}  
\text{Ch}^{^{\bullet}}(i,n)\ =\ \left(0,[H], \frac{H^{^2}}{2}-i\ell,\chi(O_{_H})-H\frac{Tod(X)}{2}-n\right).  
\end{equation}   

Semi-stability usually implies that the moduli space only deforms in the linear system. However, in most instances, \eqref{eq:mch} is not smooth. If they were, deformation of the sheaf would give the whole bundle. To circumvent this issue, we need to embed the moduli space into a smoothing

\begin{equation}   
{\cal M}\ \hookrightarrow\ {\cal A}_{_{smooth}}  
\end{equation}  

In principle, the smoothing features infinite cohomologies. However, we can always identify a map, $\phi$, such that ${\cal A}_{_{smooth}}$ is obtained from a variety with at most two cohomologies. At the level of the cotangent bundle, it reads as follows  

\begin{equation} 
\mathbb{E}^{^{\times}}\ \xrightarrow{\phi}\  \mathbb{A}_{_{smooth}}.  
\end{equation}  

Subsequently, one can take a point $p\in{\cal M}\left(X, Ch^{^{\bullet}}\right)$ such that

\begin{equation}  
\text{rank}\left(\mathbb{E}^{^x}\right)\bigg|_{_{p\in{\cal M}}}\ =\ \text{Vdim}_{_{\mathbb{C}}}\left({\cal M}\left(X, \text{Ch}^{^{\bullet}}\right)\bigg|_{_p}\right)  .   
\label{eq:rankE}  
\end{equation}   

When $X$ is Calabi-Yau, \eqref{eq:rankE} is trivial. The neighbourhood of the selected point in the singular moduli space, $R$, can therefore be plotted in terms of Ext$^{^1}(F,F)$ and Ext$^{^2}(F,F)$, defining the obstruction bundle shown in the figure below, with $F$ denoting a class associated to the point $p$   

\begin{equation}  
[F]\ =\ p\ \in{\cal M}\left(X, \text{Ch}^{^{\bullet}}\right)  
\end{equation}

\begin{figure}[ht!]    
\begin{center}
\includegraphics[scale=0.5]{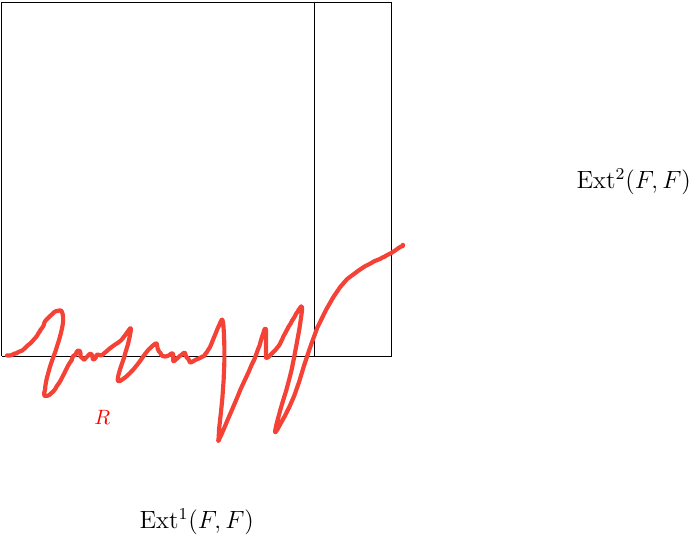}   
\caption{\small Sample sketch of an obstruction bundle for a singular CY threefold.}
\label{fig:obstrbun} 
\end{center}
\end{figure} 

To the neighbourhood of the deformation\footnote{$R$ actually denotes the set of points that are invariant under deformation.}, $R$, there is an assigned cycle, $[R]$, which, in turn can be decomposed as follows  

\begin{equation}  
[R]\ =\ \text{number}\ \cdot\ [\text{pt}],    
\label{eq:classR}   
\end{equation}  
which, as a class, should be invariant under deformation. Just to define the notation in \eqref{eq:classR}, [pt] corresponds to the point class, whereas the prefactor are nothing but the intersection numbers featuring in Gromov-Witten invariants. Setting some further notation, 

\begin{equation}   
\text{deg} [R] \ =\ [M]^{^{\text{vir}}},  
\end{equation}  
we can therefore define the DT-invariants associated to the moduli space in question, which reads as follows  

\begin{equation}  
\text{DT}\left(X, \text{Ch}^{^{\bullet}}\right)\ =\ \int_{_{[M]^{^{\text{vir}}}}}1\ =\ 1\cap[M]^{^{\text{vir}}},      
\end{equation}      
with the latter being defined by means of the generating function   

\begin{equation}   
Z_{_1}^{^H}(q)\ =\ \sum_{_n\in\mathbb{N}}\overline{\text{DT}}\left(X, \text{Ch}^{^{\bullet}}\right)q^{^n}.  
\end{equation}

\subsubsection{Hypersurface rigidity: advantages and caveats}  

The main issue encountered upon evaluating DT-invariants is ensuring transversality and stability are simultaneously satisfied.

Given a fibration  

\begin{equation}  
\begin{aligned}
H\ \hookrightarrow\ &X \\
&\downarrow\ \pi\\ 
&\mathbb{P}^{^1}  
\end{aligned}  
\label{eq:Hhyp}
\end{equation}
there is a moduli space of Hodge structures for $X$ and $H$. The latter introduces a relative moduli space of Hodge structures for the fibration 

\begin{equation}  
\begin{aligned}
&X \\
&\downarrow\\ 
&\mathbb{P}^{^1}  
\end{aligned}   
\label{eq:fibration}
\end{equation}  
corresponding to   

\begin{equation}  
p\ \in\ \mathbb{P}\ \mapsto\ H^{^2}\left(X_{_p}, K_{_{X_{_p}}}\right)  
\end{equation}

DT-invariants ultimately count only the curves surviving the deformation. \eqref{eq:Hhyp} can be extended to the case where $H\equiv K3$, in which case the partition function $Z_{_i}^{^H}(q)$ is known to be a modular form. Rigidity of the surface $H$ simplifies a lot the calculation. For example, assuming $X$ to be a quintic 3-fold in $\mathbb{CP}^{^4}$, one can take $K=1, L=H=\mathcal{O}(1)$ to be an ample divisor. Focusing only on ideal sheaves on $H$, the aim is that of making the information on the hypersurface $H$ as simple as possible. Specifically, one could look at 0-torsion sheaves, which could be realised in two different ways since they share the same Chern characters. Counting only rank-1 contributions, $K=1$. One could further assume the sheaf, $\mathcal F$, to be the pushforward of 1D schemes of $H$ in $X$.    

\begin{figure}[ht!]  
\begin{center}   
\includegraphics[scale=0.55]{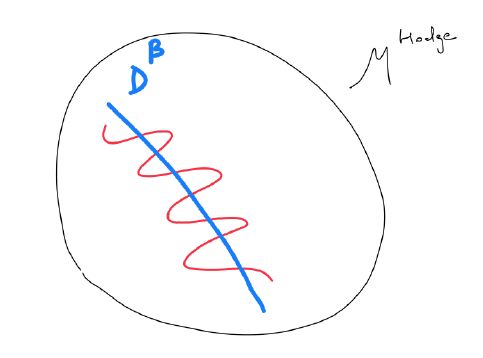} \ \ \ \ \ \ \ \ \ \ \  \ \ \ \ \ \  
\includegraphics[scale=0.5]{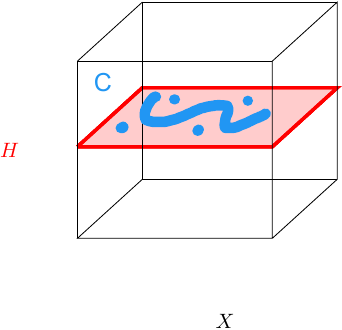}     
\label{fig:xch}  
\caption{\small The picture on the left shows a moduli space of Hodge structures. Picking a class $\beta\in H^{^2}(H,\mathbb{Z})$, $D^{^{\beta}}$ is a divisor in this analytic structure that encodes the curve (in red). On the right hand side, a curve $C$ (in blue) inside the hypersurface $H$ (in red) is taken as the union of points and a continuous curve.}
\end{center}  
\end{figure}

Take a curve and points on $H$  

\begin{equation}  
C\ =\ Z\ \cup\ P_{_1}\ \cup...\ \cup\ P_{_n}\ \subset\ H\ \subset\ X  
\label{eq:CZ}     
\end{equation} 

This induces a short-exact sequence  

\begin{equation}  
0\ \rightarrow\ \mathcal{I}_{_{C/H}}\ \rightarrow\ \mathcal{O}_{_H}\ \rightarrow\ \mathcal{O}_{_C}\ \rightarrow\ 0  
\end{equation}  
with $\mathcal{I}_{_{C/H}}$ denoting a rank-1 ideal on $H$. The Chern characters on the complement of the curve, restricted to the hypersurface, reads  

\begin{equation}  
\text{Ch}\left(\mathcal{I}_{_{C/H}}\right)\ =\ \left(1,-\beta,-n\right)  
\label{eq:chch1}
\end{equation}  

Pushing forward to $X$, \eqref{eq:chch1} reads  

\begin{equation}  
\text{Ch}\left(\iota_{_*}\mathcal{I}_{_{C/H}}\right)\ =\ \left(0, H,-\beta,-n\right)  
\label{eq:chch2}
\end{equation}
which is rank-0 on the curve, and rank-1 everywhere else. Due to the codimension-2 nature of the curve with respect to $X$, we can construct a commutative diagram between the two ideal sheaves, that further induces the following short-exact sequence   

\begin{equation}  
0\ \rightarrow\ \left[\mathcal{O}(-H)\ \rightarrow\ \mathcal{I}_{_{C/X}}\right]\ \rightarrow\ \iota_{_*}\mathcal{I}_{_{C/X}}\ \rightarrow\ 0  
\end{equation}    

Importantly, $\left[\mathcal{O}(-H)\ \rightarrow\ \mathcal{I}_{_{C/X}}\right]$ and $\iota_{_*}\mathcal{I}_{_{C/X}}$ are quasi-isomorphic to each other in the derived category, implying they deform similarly as long as  

\begin{equation}  
H^{^i}\left(\mathcal{O}_{_X}, \mathcal{I}_{_{C/H}}(H)\right)\ =\ 0\ \ \ \forall\ i>0  
\end{equation}  

But these objects are always Joyce-Song stable. Computational invariance implies that, starting from the moduli space, we have the following projective fibration  

\begin{equation}  
M\left(\mathcal{O}_{_X}(-H)\ \longrightarrow\ \mathcal{I}_{_{C/X}}\right)\ \xrightarrow{\mathbb{P}\left(H^{^{\bullet}}\left(X,\mathcal{I}_{_{C/X}}(H)\right)\right)}\ \text{Hilb}\left(C\subset X\right)  \ =\ M\left(\mathcal{I}_{_{C/X}}\right)
\label{eq:fibr}
\end{equation}

At the level of DT-invariants, \eqref{eq:fibr} leads to factorisation   as follows

\begin{equation}   
\text{DT}\left(M\left(\mathcal{O}_{_X}(-H)\ \longrightarrow\ \mathcal{I}_{_{C/X}}\right)\right)\ =\ \text{DT}\left(M\left(\mathcal{I}_{_{C/X}}\right)\right)\cdot\ \chi\left(\mathbb{P}\left(H^{^{\bullet}}\left(X,\mathcal{I}_{_{C/X}}(H)\right)\right)\right)  
\end{equation}   

But this is assuming rigidity of $H$. In particular, rigidity in this setup leads to failure in evaluating all terms in the expansion of $Z_{_i}^{^H}(q)$. Furthermore, this method features shortcomings if one were to account for deformation of subschemes.    

A further setup that one could be considering is for $H=K3$, analysed by [Ch., Sheshmani], requiring assuming Serre duality in the deformation obstruction theory. The main caveat is that allowing the surface to deform, the ideal sheaf of points stays the same, but the Cartier divisor might disappear. This is relevant in Hodge Theory, in the setting that given cohomology/homology, can we uniquely specify the corresponding variety? There might be no holomorphic cycles supporting certain curves. We therefore need to ensure that a certain deformation preserves all curves (associated to the D2-branes), not just the points (D0-branes). Need to count fibers in   \eqref{eq:fibration}, where $\beta$ after deformation can be represented by a holomorphic curve. Counting such things defines the Noether-Lefschetz number ($NL$)

\begin{equation}  
Z_{_i}^{^H}(q) =\ Z\left(\text{Hilb}^{^n}(H)\right)\cdot\ NL  
\end{equation}

With such construction, \cite{KKS} proved the S-modularity conjecture for the case of a compact CY, a surface, and sheaves supported on the surface. However, general CY3 complete intersections or quintic 3-folds, the study of S-duality requires using the degeneration technique, corresponding to taking a variety, degenerating it as follows 

\begin{equation}  
X_{_5}\ \ \ \longrightarrow\ \ \ X_{_4}\ \cup\ \mathbb{P}^{^3}  
\label{eq:ke}
\end{equation}  
where  $X_{_5}$ is the zero-locus of a function $f_{_5}$, 
\begin{equation}  
X_{_5}\ \overset{def.}{=}\ Z\left(f_{_5}\left(x_{_0},..., x_{_4}\right)\right)  
\label{eq:x5}
\end{equation}  

Then, one can construct a family that is degeneration-equal to the original setup  \eqref{eq:x5}

\begin{equation}  
tf_{_5}\left(x_{_0},...,x_{_4}\right)\ -\ f_{_1}\left(x_{_0},...,x_{_4}\right)\ +\ f_{_2}\left(x_{_0},...,x_{_4}\right)   \ \ \ \ \forall t  
\label{eq:polyn}
\end{equation}  
implying  

\begin{equation}  
Z\left(f_{_1}\right)\ =\ \mathbb{P}^{^3}\ \ \ \ \ \ ,\ \ \ \ \ \ Z\left(f_{_2}\right)\ =\ X_{_4}   
\end{equation}     

Consistency with \eqref{eq:ke} is readily proved by showing that, for $t=0$, \eqref{eq:polyn} reduces to a product of a quartic and a hyperplane, whereas for $t\neq0$ we get the original variety, $X_{_5}$.

  \eqref{eq:ke} can therefore be rewritten as two Fano varieties meeting at an anticanonical divisor under Tyurin degeneration

\begin{equation}  
X_{_5}\ \ \ \longrightarrow\ \ \ X_{_4}\ \underset{K3}{\cup}\ \mathbb{P}^{^3}  
\label{eq:ke1}
\end{equation}  
with more specific details outlined in Section \ref{sec:3.1}. However, for the formalism to apply, we need the space of degenerating family to be smooth, which is not due to the presence of a base locus. To smooth the family, one needs to perform a blow-up procedure, which is a common technique in birational geometry  

\begin{equation}  
X_{_5}\ \ \ \longrightarrow\ \ \ X_{_4}\ \cup\ B\underset{S}{|}\mathbb{P}^{^3}  
\label{eq:ke2}
\end{equation}

\subsection{Tyurin degeneration}       \label{sec:3.1}

In manifold theory, it is essential to understand how manifolds can be glued out of more elementary manifolds with boundary. From Tyurin's analysis, it is indeed known that the following setups should be treated as analogous, \cite{DCHT,KPS}:  

\begin{enumerate}   

\item   A smooth oriented $C^{^{\infty}}$-manifold $X$ with boundary $\partial X\ \subset\ X$ as encoded in the pullback map  

\begin{equation}  
C^{^{\bullet}}(X, k)\ \rightarrow\ C^{^{\bullet}}(\partial X, k)    
\end{equation}
where the boundary inherits a natural orientation  

\begin{equation}  
\text{or}:\ \Gamma\left(X, {\cal O}_{_X}\right)\ \rightarrow\ k[-d].  
\end{equation}

\item   An anticanonical divisor, $D\equiv \partial X$, in a Fano variety, $X$, and the associated pullback map on derived global sections  

\begin{equation}  
\Gamma\left(X,{\cal O}_{_X}\right)\ \rightarrow\ \Gamma\left(\partial X, {\cal O}_{_{\partial X}}\right)  
\end{equation} 
with $D$ inheriting a natural CY structure.

\end{enumerate}      

In complete analogy to gluing of oriented manifolds with boundaries, one can study a CY variety $X$ by degenerating it into the union of two Fano varieties, $Y_{_1}, Y_{_2}$, glued along a common anticanonical divisor, $S$,   

\begin{equation}  
Z:\ X\ \rightarrow\ Y_{_1}\ \underset{S}{\bigcup}\ Y_{_2}. 
\end{equation}   

\begin{figure}[ht!]    
\begin{center}
\includegraphics[scale=0.7]{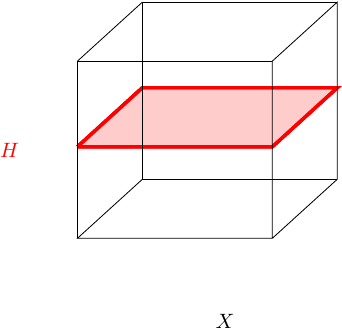}  
\ \ \ \ \ \ \ \ \ \ \ \ \ \ \ \ \ \ \ \ \ \ 
\includegraphics[scale=0.7]{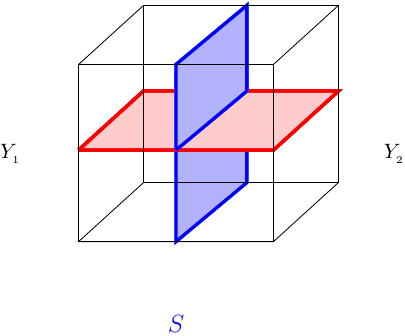}  
\caption{\small Pictorial representation of how the variety splits into two under Tyurin degeneration. The LHS is associated to a CY threefold, $X$, whereas the RHS is associated to the union of two Fano varieties, $Y_{_1}, Y_{_2}$, joined together by an anticanonical divisor, $S$.}
\label{fig:Hmodule} 
\end{center}
\end{figure} 

From the B-side perspective, this corresponds to a deformation of the category    

\begin{equation}  
\text{Perf}\left(Y_{_1}\right)\underset{\text{Perf}(X)}{\times}\text{Perf}\left(Y_{_2}\right)    
\label{eq:tyurin}   
\end{equation}      
to the category Perf$(X)$. The latter being CY, leads one to assume \eqref{eq:tyurin} to be CY as well.

\begin{equation}  
F_{_{\pm}}\ =\ \left(F_{_+}, F_{_-}\right):\ \chi_{_{2,\pm}}\ \rightarrow\ Y\ \times \ Y  
\end{equation}  
is a compatible spherical functor.

In \cite{KKS}, the authors' main focus is that of studying the moduli space of sheaves of fibers.

One would like to have factorisation  

\begin{equation}  
Z\left(\text{DT}\left(X, \text{Ch}^{^{\bullet}}\right)\right)\ =\ \sum_{_{\text{Ch}^{^{\bullet}}=\text{Ch}_{_1}^{^{\bullet}}+\text{Ch}_{_2}^{^{\bullet}}}} Z\left(\text{DT}\left(Y_{_1}/S, \text{Ch}_{_1}^{^{\bullet}}\right)\right)Z\left(\text{DT}\left(Y_{_2}/S, \text{Ch}_{_2}^{^{\bullet}}\right)\right).  
\label{eq:factorisation}
\end{equation} 

The main issue with \eqref{eq:factorisation} is the fact that it does not ensure transversality. Furthermore, whenever performing calculations of such invariants, we must always bear in mind the delicate balance in between stability and transversality. Oftentimes, the one is traded for the other. However, for the calculation not to miss relevant information due to the factorisation of the moduli space, both properties need to be satisfied simultaneously.  

The calculation of such partition function is related to that of BPS BH microstate counting for a D4-D2-D0-brane system wrapping the $H$-surface in $X_{_{\text{CY3}}}$, as originally addressed by Aganagic, Ooguri, Saulina, and Vafa [2005], where the latter showed the following  

\begin{equation}  
\begin{aligned}
Z_{_{BH}}\ &=\ \left|Z_{_{top}}\right|^{^2} \\
&=\int\prod_{_i=1}^{^{|2g-2|}}dU_{_i}\left|Z_{_{\text{top}}}\left(U_{_1},...,U_{_{|2g-2|}}\right)\right|^{^2},    
\end{aligned}
\end{equation}   
where $U_{_i}$ defines the holonomy of the gauge field turned on in the $i^{^{th}}$ D-brane stack.

\subsubsection{Sheave splitting}

Any sheaf $\mathcal{F}\in X$ or $\mathcal{F}_{_i}\in Y_{_i}$, $i=1,2\in\text{Coh}(\mathbb{P})$. There exists a short-exact-sequence  

\begin{equation}  
0\ \rightarrow\ \mathcal{F}\ \rightarrow\ \mathcal{F}_{_1}\oplus\mathcal{F}_{_2}\  \rightarrow\ \mathcal{F}_{_1}\ \overset{\iota}{\otimes}\ \mathcal{O}_{_S}\ \simeq\ \mathcal{F}_{_2}\overset{\iota}{\otimes}\mathcal{O}_{_S}\ \rightarrow\ 0  
\end{equation}  

This is where the issue arises: tensoring a codimension-1 sheaf with something that is also codimension-1. This is not the usual tensor product, in particular, it does not lie in $D^{^b}\text{Coh}(X)$. From this, it thereby follows that the $\mathcal{F}_{_i}$ sheaves are not necessarily meeting $S$ transversally, in particular the relative setting features a nontrivial torsion element  

\begin{equation}  
\text{Tor}_{_1}^{^{\mathcal{O}_{_{Y_{_i}}}}}\left(\mathcal{F}_{_i},\mathcal{O}_{_S}\right)\ \neq \ 0  
\end{equation}   

Requiring compactification to make sheaves transverse  

\begin{equation}  
M\left(Y_{_i}/S,\text{Ch}_{_i}^{^*}\right)\ \Subset\ M\left(Y_{_i},\text{Ch}_{_i}^{^*}\right)
\end{equation}   

Alternatively to the blow-up, one can perform acordionical expansion, then gluing.  Correspondingly, the curve \eqref{eq:CZ} splits into two parts (with respect to the relative moduli spaces)

\begin{equation}  
M(C\hookrightarrow X)\ \rightarrow\ \overline{M\left(C_{_1}\hookrightarrow Y_{_1}/S\right)}\ \cup\ \overline{M\left(C_{_2}\hookrightarrow Y_{_2}/S\right)}
\end{equation}    
with respective virtual classes  

\begin{equation}  
\left[\bar M\left(C_{_1}\hookrightarrow Y_{_1}/S\right)\right]\ \ \ \ \ \ , \ \ \ \ \ \ \left[\bar M\left(C_{_2}\hookrightarrow Y_{_2}/S\right)\right]  
\end{equation}  

\begin{equation}   
\left[M_{_1}\underset{\text{Hilb}^{^n}}{\times}M_{_2}\right]^{^{\text{vir}}}\ \overset{def.}{=}\ \Delta^{^!}\left(\left[M_{_1}\right]^{^{\text{vir}}}\times\left[M_{_2}\right]^{^{\text{vir}}}\right) \ \equiv\ \left[M\left(C\hookrightarrow\ X\right)\right]^{^{\text{vir}}}
\end{equation}
with the last equality following by deformation invariance.

When it is fully transversal, the Hilbert polynomial factorises 

\begin{equation}  
\text{Hilb}^{^{n_{_1}}}(S)\times\text{Hilb}^{^{n_{_2}}}(S)
\end{equation} 

Upon further relaxing the assumption of the sheaves being ideal, the degeneracy technique fails in its ordinary setup, and one therefore needs to resort to derived intersection theory, instead, \cite{KKS}. Considering the moduli space for DG-schemes of rigidified perfect complexes, one can achieve a suitably upgraded counterpart to \eqref{eq:ke1}, on which the shifted symplectic structure formalism can still be applied as explained in Section \ref{sec:1}.

For the purpose of our work, the main feature of the result by \cite{KKS} is that the degeneracy formula in presence of non-ideal sheaves, can be achieved and relates two DG-categories that are related by deformation quantisation. However, as will be explained in Section \ref{sec:5}, this is not yet the most general setup that one could consider. Specifically, it still falls within the abelianised setup.

\subsection{A-model relative CY structure}     \label{sec:3.2}

To explain the terminology, we refer to the following Theorems stated and proved in \cite{KPS}  

\subsection*{Relative right CY structures}  

Let ${\cal Y}$ be a locally proper category, and  

\begin{equation}  
\Xi:\ HH_{_{\bullet}}({\cal Y})^{^{\text{V}}}\ \rightarrow\ \text{Map}_{_{{\cal X}\otimes{\cal X}^{^{\text{op}}}}}\left({\cal Y}_{_{\Delta}},{\cal Y}^{^{\text{V}}}\right)  
\end{equation} 
the natural morphism 

\begin{equation}  
HH_{_{\bullet}}({\cal Y})\ \simeq\ {\cal Y}_{_{\Delta}}\underset{{\cal Y}\otimes{\cal Y}^{^{\text{op}}}}{\otimes}{\cal Y}_{_{\Delta}}.   
\end{equation}  

Then, a \emph{weak} $d$-dimensional right CY structure is a map  

\begin{equation}   
\phi:\ HH_{_{\bullet}}({\cal Y})\ \rightarrow\ \mathbb{k}[-d]  
\end{equation}  
such that

\begin{equation}  
\Xi(\phi):\ {\cal Y}_{_{\Delta}}[d]\ \rightarrow\ {\cal Y}^{^{\text{V}}}  
\end{equation} 
is a weak equivalence.   

A $d$-dimensional right CY structure on ${\cal Y}$, instead, is an $S^{^1}$-equivariant map     

\begin{equation}   
\tilde\phi:\ HH_{_{\bullet}}({\cal Y})_{_{S^{^1}}}\ \rightarrow\ \mathbb{k}[-d]  
\end{equation}   
such that the composite map    

\begin{equation}   
HH_{_{\bullet}}({\cal Y})_{_{S^{^1}}}\ \rightarrow\ HH_{_{\bullet}}({\cal Y})_{_{S^{^1}}}\ \rightarrow\ \mathbb{k}[-d]  
\end{equation}  
is a weak CY structure.  

A functor $F:\ {\cal X}\rightarrow {\cal Y}$ between stable $\infty$-categories is spherical if:  

\begin{itemize}  

\item   $F$ admits a left and a right adjoint, $F^{^*}, F^{^!}$, respectively.

\item  The homotopy cofiber $T_{_{\cal Y}}$ of the counit   $F\circ F^{^!}\ \rightarrow\ \text{id}_{_{\cal Y}}$ is an autoequivalence of ${\cal Y}$.

\item  The homotopy fiber $T_{_{\cal X}}$ of the unit $\text{id}_{_{\cal X}}\ \rightarrow\ F^{^!}\circ F$ is an autoequivalence of ${\cal X}$.

\end{itemize}    

\medskip 

\medskip 

\underline{Theorem 1.18 [KPS]}  Let ${\cal Y}$ be a CY category, and ley 

\begin{equation}  
F:\ {\cal X}\ \rightarrow\ {\cal Y}   
\end{equation}
be a functor that admits left and right adjoints. Then, $F$ has a weak relative right CY structure if and only if it has the structure of a compatible spherical functor.

\medskip 

\medskip 

\underline{Theorem 1.19 [KPS]}     Let 

\begin{equation} 
{\cal W}:\ X\ \rightarrow\ \mathbb{C}  
\end{equation}
be an admissible LG model with generic fiber $Y$. Then, the cap functor  

\begin{equation}  
\cap:\ \text{FS}(X,{\cal W})\ \rightarrow\ \text{Fuk}({\cal Y})  
\label{eq:capfunctor}   
\end{equation}   
carries a natural relative CY structure\footnote{Theorems 1.18 and 1.19, \cite{KPS}, imply that $\cap$ is spherical.}. \eqref{eq:capfunctor} sends a Lagrangian $L$ in $X$ to $L\cap X_{_t}$, with $X_{_t}\equiv Y$, and $t\in\mathbb{C}-\{p_{_i}\}$, $\{p_{_i}\}$ denoting the critical values of ${\cal W}$.

\subsection*{A-model relative CY structure}

Given two functors  

\begin{equation}  
\cap\ :\ \text{FS}(X,{\cal W})\ \rightarrow\ \text{Fuk}(Y)  
\ \ \ \ \ \ \ \ , \ \ \ \ \ \ \ \ 
\cup\ :\ \text{Fuk}(Y)\ \rightarrow\ \text{FS}(X,{\cal W}),   
\end{equation}    
with the former defining the restriction functor to the fiber at $\infty$, and the latter being the Orlov functor\footnote{The Orlov functor corresponds to a Hamiltonian flow of a Lagrangian along an arc from +$\infty$ to itself going around all the critical values.}. According to the HMS dictionary\footnote{Succinctly overviewed in appendix \ref{sec:A}.} the mirror of $(X,{\cal W})$ is a Fano variety, $X^{^{\text{V}}}$, and the categorical equivalence in question thereby reads  

\begin{equation}    
\text{FS}(X,{\cal W})\ \simeq\ \text{Perf}\left(X^{^{\text{V}}}\right).   
\end{equation}   

Under HMS, the restriction functor, $\cap$, turns into 

\begin{equation}  
\cap\ :\ \text{Perf}\left(X^{^{\text{V}}}\right)\ \rightarrow\ \text{Fuk}\left(Y^{^{\text{V}}}\right),       
\end{equation}    
from $X^{^{\text{V}}}$ to an anticanonical divisor

\begin{equation}  
Y^{^{\text{V}}}\ \hookrightarrow\ X^{^{\text{V}}}.          
\end{equation}

Among the original motivations that led \cite{KPS} to develop the formalism of spherical functors in relation to relative CY structures, \cite{KPS}, is the flexibility provided by the noncommutative framework for bringing further insight into classical hyperk$\ddot{\text{a}}$hler geometry by generalising to derived stacks.

\section{Perturbative Topological String Theory}  \label{sec:3} 

We now start relating the tools outlined in the first sections to more theoretical physics applications. In doing so, we skip most of the intermediate steps bridging the two languages, and we refer the interested reader to the respective original literature in this regard, mostly \cite{Pan, AOSV}. Comparison in between the mathematics and theoretical physics sources will follow through accordingly. 

The structure of this section is as follows: 

\begin{enumerate} 

\item   Section \ref{sec:4.1} the generating function calculated at the end of Section \ref{sec:2} with that of BPS black holes in D4-D2-D0-branes systems, following the work of \cite{Pan, AOSV}.

\item   Section \ref{sec:4.2} provides a TFT formulation of gluing rules, practically leading to the formulation of Gromov-Witten (GW)-invariants in terms of cobordisms equipped with complex line bundles.

\end{enumerate}   

\subsection{Black holes and CY $X=L_{_1}\oplus L_{_2}\rightarrow\Sigma_{_g}$}  \label{sec:4.1}

Consider type-IIA string theory computed on a CY manifold, $X$, 

\begin{equation}  
L_{_1}\oplus L_{_2}\rightarrow\Sigma_{_g}  
\end{equation}  
subject to the constraint  

\begin{equation}  
\text{deg}(L_{_1})\ +\ \text{deg}(L_{_1})\ =\ -\chi\left(\Sigma_{_g}\right). 
\end{equation}

 In the context of topological string theory, the aim is that of counting bound states of D4-D2-D0-branes, where the D4-branes wrap  

\begin{equation}  
C_{_4}\ =\ L_{_2}\ \rightarrow\ \Sigma_{_g}  
\end{equation}  
and D2-branes wrap $\Sigma_{_g}$.

As shown by \cite{AOSV}, the black hole partition function for type-II superstring theory on a CY threefold, reads

\begin{equation} 
\begin{aligned}
Z_{_{BH}}\ &=\ \bigg|Z_{_{\text{TOP}}}\bigg|^{^2}   \\
&=\int\prod_{_{i=1}}^{^{|2g-2|}}dU_{_i}\bigg|Z_{_{\text{TOP}}}\left(U_{_1},...,U_{_{|2g-2|}}\right)\bigg|^{^2},   
\label{eq:partfn}      
\end{aligned}
\end{equation}    
with $U_{_i}$ denoting the holonomy of the gauge field in the $i^{th}$-stack of D-branes.     

The D4-D2-D0-brane system is a configuration that can be realised by a suitable choice of a curve $C$ on $H\subset X$. In Section \ref{sec:2}, we highlighted the importance of assuming such curve to encode all the information of the CY, $X$, to calculate the corresponding DT-invariant. Likewise, the same thing can be stated for the calculation of the topological partition function, \eqref{eq:partfn}, where all the information is assumed to be contained in the specified brane system.

\subsection{TFT formulation of gluing rules}    \label{sec:4.2}

The gluing of the residue theory of rank-2 bundles on curves is most concisely formulated as a functor of tensor categories  

\begin{equation}  
\text{GW}(-):\ 2\text{Cob}^{^{L_{_1},L_{_2}}}\ \rightarrow\ \text{R-mod}.  
\label{eq:GW}   
\end{equation}  

As the notation suggests, \eqref{eq:GW} provides a functorial formulation of Gromov-Witten invariants in terms of cobordisms equipped with complex line bundles, $L_{_1},L_{_2}$. Deformation-invariance of the residue allows for a topological formulation of the gluing structure.

\begin{figure}[ht!]    
\begin{center}
\includegraphics[scale=0.35]{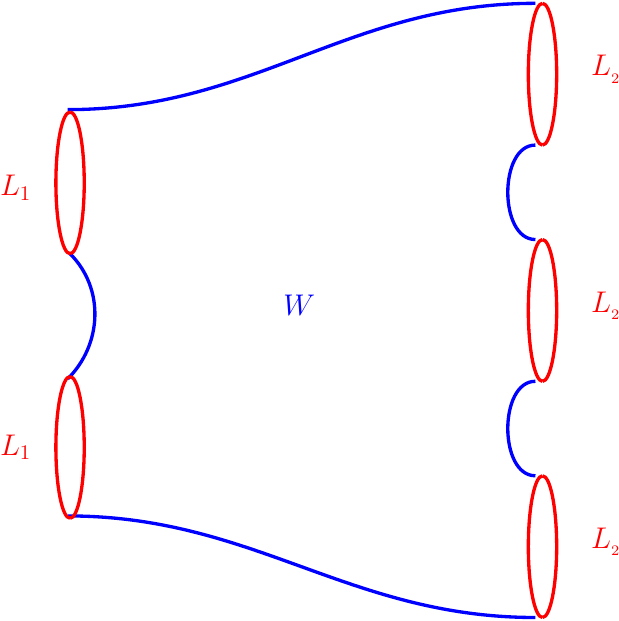}   
\caption{\small An example of a cobordism on which to calculate the topological string partition function.}
\label{fig:wl1l2} 
\end{center}
\end{figure} 

\begin{figure}[ht!]    
\begin{center}
\includegraphics[scale=1]{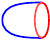}   \ \ \ \ \ \ \ \ \ \ \ \ \ \ \ \ \ \ \ \includegraphics[scale=0.5]{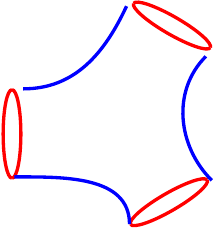}  
\caption{\small Fundamental constitutents on which gluing rules can be applied to recover an arbitrary topological string amplitude.}
\label{fig:cap} 
\end{center}
\end{figure} 

Under the assumption of a stable smooth variety, the amplitude on any cobordims can be recast to the gluing of fundamental constituents, namely the CY cap, $C^{^{(-1,0)}}$, and the CY pair of pants, $P^{^{(1,0)}}$, (cf. Figure \ref{fig:wl1l2}), whose respective amplitudes read  

\begin{equation}  
Z^{^{\text{top}}}\left(C^{^{(-1,0)}}\right)\ =\ \sum_{_R}d_{_q}(R)\ q^{^{-k_{_R}/4}}\text{Tr}_{_R}U  \ \ \ \ \ \ \ \ \ \ \ ,\ \ \ \ \ \ \ \ \ \ U\ \overset{def.}{=}\ {\cal P}\ e^{^{\oint A}}  
\label{eq:cap}   
\end{equation} 
with $U$ denoting the holonomy of the gauge field on the D-branes around a circle where the D-branes meet $\Sigma$. Taking the number of D-branes to be infinite, $R$ labels a representation of $SU(\infty)$, and 

\begin{equation}  
q\ \overset{def.}{=}\ e^{^{-g_{_s}}}.   
\end{equation}

The coefficient $d_{_q}(R)$ is the quantum dimension of the symmetric group representation corresponding to the Young Tableaux of $R$, defined as follows  

\begin{equation}  
d_{_q}(R)\ =\ \prod_{_{\square\in R}}\frac{1}{[h(\square)]_{_q}}.   
\end{equation}   

On the other hand, the pair of pants come equipped with three sets of representations, corresponding to 3 stacks of D-branes at three semi-infinite tubes with holonomies $U_{_i}$, $i=1,2,3$, thereby leading to the corresponding amplitude

\begin{equation}  
Z^{^{\text{top}}}\left(P^{^{(1,0)}}\right)\ =\    \sum_{_R}\frac{1}{d_{_q}(R)}\ q^{^{k_{_R}/4}}\text{Tr}_{_R}U_{_1}\ \text{Tr}_{_R}U_{_2}\ \text{Tr}_{_R}U_{_3}.   
\label{eq:pairofpants}   
\end{equation}   

\eqref{eq:cap} and \eqref{eq:pairofpants}, together with 

\begin{equation}  
Z^{^{\text{top}}}\left(C^{^{(0,-1)}}\right)\ =\ \sum_{_R}d_{_q}(R)\ q^{^{k_{_R}/4}}\text{Tr}_{_R}U     
\label{eq:cap1}   
\end{equation} 

\begin{equation}  
Z^{^{\text{top}}}\left(P^{^{(0,1)}}\right)\ =\    \sum_{_R}\frac{1}{d_{_q}(R)}\ q^{^{-k_{_R}/4}}\text{Tr}_{_R}U_{_1}\ \text{Tr}_{_R}U_{_2}\ \text{Tr}_{_R}U_{_3}.   
\label{eq:pairofpants1}   
\end{equation}   

Still under the assumption that the variety is smooth, an arbitrary pair of amplitudes, $\Sigma_{_L},\Sigma_{_R}$, built in terms of the fundamental constituents \eqref{eq:cap}, \eqref{eq:pairofpants}, \eqref{eq:cap1}, \eqref{eq:pairofpants1}, can, in turn, be glued together, by inverting the holonomy in one of them  

\begin{equation}  
U\ \rightarrow\ U^{^{-1}}  
\end{equation}  
ultimately reading 

\begin{equation}  
Z^{^{\text{top}}}\left(\Sigma_{_{L\cup R}}\right)\ =\   \int d U  Z^{^{\text{top}}}\left(\Sigma_{_L}\right)(U) Z^{^{\text{top}}}\left(\Sigma_{_R}\right)  \left(U^{^{-1}}\right)  \ \ \ \ \ \ , \ \ \ \ \ \  Z^{^{\text{top}}}\left(\Sigma\right) \ =\ \sum_{_R}\tilde Z^{^{\text{top}}}_{_R}\text{Tr}_{_R}(U)  
\end{equation} 
following from orthogonality of the characters  

\begin{equation}  
\int dU\ \text{Tr}_{R_{_1}}U\ \text{Tr}_{R_{_1}}U^{^{-1}}\ =\ \delta_{_{R_{_1}R_{_2}}}.     
\end{equation}

Figure \ref{fig:wl1l2} corresponds to critical points of the systole function. But, in principle, we should explore beyond the critical points, namely considering considering configurations at arbitrary points of the flow, interpolating between different SOTs.

\section{Cohomological Field Theories (CohFTs)}   \label{sec:4}

The present section explains the main setup of interest for our analysis. Specifically, Cohomological Filed Theories (CohFTs). The reason for this is that they are useful in, both, HMS and String Theory calculations. The former is well known since the pioneering work of Kontsevich in this regard. As for the latter, Moore and Segal, \cite{MS}, started addressing the more physical side of the formalism, that later developed in the work of Moore and Tachikawa, \cite{MT}. In what follows, the present section briefly overviewes the main tools of the two perspectives, highlighting how they come together in the specific context of quiver varieties describing $A_{_n}$-singularities. 

Combining the topics covered in the first part of our treatment, with HMS in mind, the present section is structured as follows: 

\begin{enumerate}  

\item   At first (in Section \ref{sec:5.1}), we introduce the notion of Cohomological Field Theories (CohFTs) as a bridging language between String Theory and HMS, most clearly arising in the formulation of the WDVV equation on Kontsevich's moduli space of stable maps.

\item   We then turn to a particular realisation of quantum cohomology in K-theory, \ref{sec:6.2}. The reason being the analogous treatment performed by Moore and Segal, \cite{MS}, and its applicability to the categorification of D-branes in String Theory. Section \ref{sec:5.3} constitutes a specific application of the formalism initiated by \cite{MT}, also known as the Moore-Tachikawa varieties. 

\item  The concluding part of this section, \ref{sec:5.4}, explains the role of Coulomb branches towards classifying 2D TFTs, and, in turn, CohFTs. The importance of extending this within the context of hyperk$\ddot{\text{a}}$hler quotients will be highlighted.

\end{enumerate}

\subsection*{CohFTs}

Initially introduced by Kontsevich and Manin, CohFTs feature as starting point a triple $(V, \eta,\mathbf{1})$, where

\begin{itemize}  

\item  $V$ is a finite-dimensional $\mathbb{Q}$-vector space  

\item $\eta$ ia a non-degenerate symmetric 2-form on $V$  

\item  $\mathbf{1}\in V$ is a distinguished element.  

\end{itemize} 

Given a $\mathbf{Q}$-basis, $\{e_{_i}\}$ of $V$, the symmetric form $\eta$ can be written as a matrix 

\begin{equation}  
\eta_{_{jk}}\ =\ \eta(e_{_j},e_{_k}).  
\end{equation}

A CohFT, then consists of a system 

\begin{equation} 
\Omega\ =\ (\Omega_{_{g,n}})_{_{2g-2+n>0}}  
\end{equation}   
of tensors  

\begin{equation}  
\Omega_{_{g,n}}\ \in\ H^{^{\bullet}}\left(\overline{\mathfrak{M}}_{_{g,n}},\mathbb{Q}\right)\otimes(V^{\bullet})^{\otimes n},  
\end{equation}   
associating a cohomology in $H^{^{\bullet}}\left(\overline{\mathfrak{M}}_{_{g,n}},\mathbb{Q}\right)$ to vectors $v_{_1}, ..., v_{_n}\in V$ assigned to the $n$ markings.

\subsection*{Kontsevich's moduli space of stable maps}

In the context of HMS, the WDVV equations are particularly useful for calculations involving the A-side, as explained in what follows. To define Gromov-Witten invariants of a symplectic manifold, ${\cal M}$, consider moduli spaces of stable homolorphic maps, \cite{Pan},  

\begin{equation}  
\overline{\cal M}_{_{g,n,\beta}}(M)\ \overset{def.}{=}\ \left\{u:\Sigma_{_g}\rightarrow X \ \text{holomorphic},\  u[\Sigma_{_g}]=\beta\right\}\ / \sim,  
\end{equation}  
there are maps  

\begin{equation}  
\overline{\cal M}_{_{g,n}}\ \longleftarrow\ \overline{\cal M}_{_{g,n,\beta}}\ \longrightarrow\ M^{^n}  
\end{equation}  

\begin{equation}  
\Rightarrow \ \ \ I_{_{g,n,\beta}}:\ \ H^{^{\bullet}}(M)^{^{\otimes n}}\ \rightarrow\ H^{^{\bullet}}\left(\overline{\cal M}_{_{g,n,\beta}}\right).    
\end{equation}

Then, the maps  

\begin{equation}  
I_{_{g,n}}\ =\ \sum_{_{\beta\in H_{_2}(M)}}r^{^{\omega(\beta)}}\ I_{_{g,n,\beta}}   
\end{equation}  
form a $\mathbb{C}[[r]]$-linear cohomological field theory.  

The Fukaya category should fit in an open-closed CohFT with the GW invariants. Kontsevich conjectured that (in good cases) the whole CohFT structure of the GW invariants can be reconstructed from the Fukaya category by taking Hochschild cohomology: 

\begin{enumerate}  

\item There should be a natural CohFT structure on the Hochschild cohomology of an $A_{_{\infty}}$ category (Deligne conjecture). 

\item There should be an isomorphism  

\begin{equation}  
H^{^{\bullet}}(M)\ \simeq\ HH^{^{\bullet}}({\cal F}(M))  
\end{equation}  
as $\mathbb{C}[[r]]$-linear CohFTs.  

\end{enumerate}    

The open-closed CohFT picture is closely related to the open-closed TFT setup addressed by Moore and Segal, \cite{MS}, in attempting to develop a mathematical formulation for D-branes arising in String Theory. 

\subsection{WDVV equations}   \label{sec:5.1}

The WDVV equation is usually defined in term of the generating function for correlators   

\begin{equation}  
F(t,Q)\ =\ \sum_{_d}\sum_{_{n=0}}^{^{\infty}}\frac{Q^{^d}}{n!}(t,...,t)_{_{n,d}},    
\end{equation}
where $d\in H_{_2}(X,\mathbb{Z})$ runs in the Mori cone of degrees, and the correlators  

\begin{equation} 
(\phi_{_1},...\phi_{_n})_{_{n,d}} 
\end{equation}  
are defined using the evaluation maps at the marked points   

\begin{equation}  
\text{ev}_{_1}\times...\times\text{ev}_{_n}:\ X_{_{n,d}}\ \rightarrow\ X\times...\times X.   
\end{equation}

In the cohomological theory, we can pull the $n$-cohomology classes of $X$, $\phi_{_1},...,\phi_{_n}\in H^{^{\bullet}}(X,\mathbb{Q})$, back to the moduli space, $X_{_{n,d}}$, to define the correlator as follows  

\begin{equation}  
(\phi_{_1},...\phi_{_n})_{_{n,d}} \ \overset{def.}{=}\ \int_{_{[X_{n,d}]}}\text{ev}_{_1}(\phi_{_1})\wedge...\wedge\text{ev}_{_n}(\phi_{_n}).   
\end{equation} 

Let $\{\phi_{_{\alpha}}\}$ denote a graded basis in $H^{^{\bullet}}(X,\mathbb{Q})$, and  

\begin{equation}  
g_{_{\alpha,\beta}}\ =\ \left<\phi_{_{\alpha}},\phi_{_{\beta}}\right>\ =\ \int_{_{[X]}}\phi_{_{\alpha}}\wedge\phi_{_{\beta}}   
\end{equation}  
be the intersection matrix.   

Then, the quantum cup product   

\begin{equation}  
\left<\phi_{_{\alpha}}\bullet\phi_{_{\beta}},\phi_{_{\gamma}}\right>\ \overset{def.}{=}\ F_{_{\alpha\beta\gamma}}(t) 
\end{equation}  
satisfies associativity relations corresponding to the WDVV equation. More generally, we can define  

\begin{equation}  
S_{_{\alpha,\beta}}\ =\ g_{_{\alpha,\beta}}\ +\ \sum_{_d}\sum_{_{n=0}}^{^{\infty}}\frac{Q^{^d}}{n!}\left(\phi_{_{\alpha}},t,...,t, \frac{\phi_{_{\beta}}}{1-qL}\right)_{_{n,d}},      
\end{equation}    
now being the solution to the following linear PDE system  

\begin{equation}  
\left(1-q\right)\partial_{_{\alpha}} S\ =\ \left(\phi_{_{\alpha}}\bullet\right)\ S.    
\end{equation}  

For the case that $X$ is a flag variety, $S$ is referred to as the \emph{Whittaker function}.

\begin{figure}[ht!]    
\ \ \ \ \ \ \ \ \ \ \ \ \ \ \ \ \ \ \ \ \ \ \ \ \ \ \ \ \includegraphics[scale=0.5]{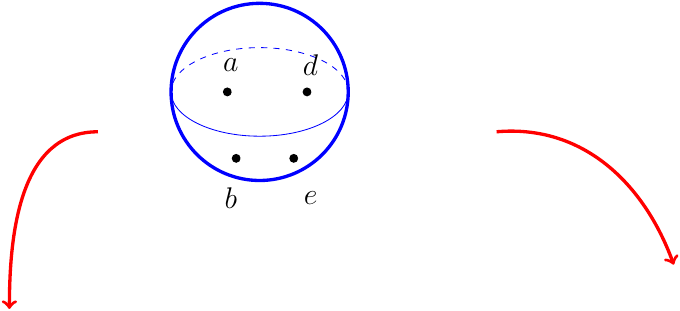}\\
\includegraphics[scale=0.5]{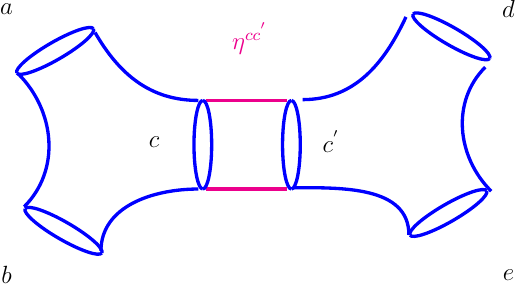}   \ \ \ \ \ \ \ \ \ \ \ \ \ \ \ \ \ \ \ \ \ \ \ \ \ \ \ \ \ \ \ \ \ \ 
\includegraphics[scale=0.5]{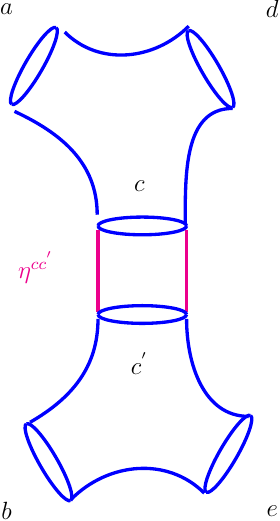}  
\caption{\small This figure shows the possible gluings leading to a 4-punctured sphere. As such, each possible decomposition is nothing but a term featuring in the WDVV equation, \eqref{eq:WDVV}.}     
\label{fig:wdvv} 
\end{figure} 

Referring to Figure \ref{fig:wdvv}, the WDVV equation reads

\begin{equation}   
\boxed{\ \ \ \ \frac{\partial^{^3}{\cal F}}{\partial T_{_a}\partial T_{_b}\partial T_{_c}}\ \eta^{cc^{^{\prime}}}\ \frac{\partial^{^3}{\cal F}}{\partial T_{_e}\partial T_{_d}\partial T_{_{c^{^{\prime}}}}}\ \ =\ (b\longleftrightarrow d) \color{white}\bigg]\color{black}\ \ },    
\label{eq:WDVV}   
\end{equation}  
where the generating function for rational curves passing through cycles $C_{_1},...,C_{_N}$ reads  

\begin{equation}  
{\cal F}(T)\ \overset{def.}{=}\ \sum_{_d}q^{^d}\mathfrak{N}_{_d}^{^X}\left(\frac{C(T),...,C(T)}{k!}\right)  
\end{equation}  
for a universal cycle defined as follows

\begin{equation}   
C(T)\ \overset{def.}{=}\ \sum_{_a}T^{^a}C_{_a}.    
\end{equation}  

\eqref{eq:WDVV} is essentially the starting point to the original formulation of Gromov-Witten theory.

One of the main remarks of the present work is that \eqref{eq:WDVV} is equivalent to the following isomorphism, \cite{KPS},  

\begin{equation}  
\boxed{\ \ \ \text{Map}_{_{{\cal Y}\otimes{\cal Y}^{op}}}\left({\cal Y}_{_{\Delta}}, {\cal Y}^{^{\text{V}}}\right)\ \simeq\ \text{Map}_{_{\text{Fun}({\cal Y},{\cal Y})}}\left(\text{id}_{_{\cal Y}}, S_{_{{\cal Y}}}\right)\color{white}\bigg]\color{black} \ \ }.
\label{eq:KPS}
\end{equation}

The original derivation of \eqref{eq:KPS}  in \cite{KPS} arises  in the context of spherical functors and relative CY structures introduced\footnote{Recalling the terminology introduced in Secion \ref{sec:3.2}.} in Section \ref{sec:3.2}, and goes as follows, \cite{KPS}.

The diagonal bimodule 

\begin{equation}  
{\cal Y}_{_{\Delta}}:\ {\cal Y}^{^{\text{op}}}\otimes{\cal Y}\ \rightarrow\ \text{Mod}_{_{\mathbb{k}}}   
\end{equation}  
with  

\begin{equation}  
{\cal Y}_{_{\Delta}}(x,y)\ \overset{def.}{=}\ \text{Map}_{_{\cal Y}}(x,y)\ \ \ , \ \ \ \forall x,y\ \in\ {\cal Y}.     
\end{equation}    

\begin{equation}  
{\cal Y}^{^{\text{V}}}:\ {\cal Y}\otimes{\cal Y}^{^{\text{op}}}\ \rightarrow\ \text{Mod}_{_{\mathbb{k}}}   
\end{equation} 
with  

\begin{equation}  
{\cal Y}^{^{\text{V}}}(x,y)\ \overset{def.}{=}\ \text{Map}_{_{\cal Y}}(x,y)^{^{\text{V}}}\ =\ \text{Map}_{_{\cal Y}}\left(\text{Map}_{_{\cal Y}}(x,y), \mathbb{k}\right).     
\end{equation}    

If ${\cal Y}$ is locally proper, then a functor representing a bimodule ${\cal Y}^{^{\text{V}}}$ is equivalent to a Serre functor, $S_{_{\cal Y}}$,   

\begin{equation}  
{\cal Y}^{^{\text{V}}}(-,-)\ \simeq\ \text{Map}\left(-,S_{_{\cal Y}}(-)\right).  
\end{equation}   

From this and the definition of ${\cal Y}^{^{\text{V}}}$ it therefore follows that   

\begin{equation}  
\text{Map}_{_{\cal Y}}(x,y)^{^{\text{V}}}\ =\ \text{Map}\left(y,S_{_{\cal Y}}(y)\right),        
\end{equation}   
and a (weak) right CY structure, $\phi$, is equivalent to the data of the equivalence of functors

\begin{equation}  
\text{id}_{_{\cal Y}}[d]\ \simeq\ S_{_{\cal Y}}  
\end{equation}  
corresponding to $\Xi(\phi)$ under the identification map  

\begin{equation}  
\text{Map}_{_{{\cal Y}\otimes{\cal Y}^{^{\text{V}}}}}({\cal Y}_{_{\Delta}},{\cal Y}^{^{\text{V}}})\ \simeq\ \text{Map}_{_{\text{Fun}({\cal Y},{\cal Y})}}\left(\text{id}_{_{\cal Y}},S_{_{\cal Y}}\right).            
\end{equation}

\subsection{Moore-Tachikawa varieties}  \label{sec:5.3}

As already mentioned, CohFTs are particularly useful for the categorification of D-branes in String Theory, \cite{MS, MT}. The present subsection briefly overviews the setup of \cite{MT}, explaining its reltion with the first sections of the present treatment. In particular, we emphasise that, unlike the original works of \cite{MS, MT}, our work is motivated by advancements in HMS towards addressing String Theory questions.

Quiver varieties provide a very useful setup where to study HMS. For the purpose of what follows, let us first briefly outline the main points of interest for the present work. 

A quiver variety is defined by a group action   

\begin{equation}  
GL(V)=\prod_{i}GL\left(V_{_i}\right)\ \circlearrowright\  \mathcal{M}({\mathcal Q, V, W}),
\end{equation}
and a moment map  

\begin{equation}  
\mu: \mathcal{M}({\mathcal Q, V, W})\ \rightarrow\ \mathfrak{gl}(V)^{^*}. 
\label{eq:4}
\end{equation}

From \eqref{eq:4}, one can define the Nakajima quiver varieties  

\begin{equation}  
\mathfrak{M}_{_{\theta}}\ \overset{def.}{=}\ \mu^{^{-1}}(0)\ //_{_{\theta}} GL(V),
\label{eq:Mt}
\end{equation}
with $\theta$ denoting the choice of a character. In Theoretical Physics terms, \eqref{eq:Mt} is also referred to as the Higgs branch. By construction, $\mathfrak{M}_{_{\theta}}$ is a smooth algebraic variety, with \eqref{eq:Mt} usually being referred to as the GIT quotient. When $\theta=0$, the corresponding Nakajima quiver variety,  

\begin{equation}  
\mathfrak{M}_{_{0}}\ \overset{def.}{=}\ \mu^{^{-1}}(0)\ // GL(V)
\label{eq:Mo}
\end{equation}
is affine singular. Nakajima proved how to relate \eqref{eq:Mt} and \eqref{eq:Mo} by means of a conic symplectic resolution (CSR)

\begin{equation}  
\pi:\ \mathfrak{M}_{_{\theta}}\ \longrightarrow\ \mathfrak{M}_{_{0}}.
\end{equation}

A Theorem asserts that given a CSR, its kernel are Lagrangian submanifolds

\begin{equation}  
{\cal{L}}\ \overset{def.}{=}\ \pi^{^{-1}}(0)\ =\ \bigcup_{_{\alpha}}\ \cal{L}_{\alpha}.
\end{equation}

4D ${\cal N}=2$ SCFTs admit the following functorial prescription, \cite{MT}, 

\begin{equation}  
\eta_{_{G_{_{\mathbb{C}}}}}:\ \text{Bo}_{_2}\ \longrightarrow\ {\cal C}  
\label{eq:MT}    
\end{equation}  
as a 2D TFT, with source and target being 2-categories. Bo$_{_2}$ denotes the bordism 2-category, whereas ${\cal C}$ deontes a symplectic manifold. For the purpose of our work, \eqref{eq:MT} fits as a CohFT in the context of HMS. The original work of Moore and Tachikawa was mostly focused on type $A_{_n}$-singularities. 

Objects of Bo$_{_2}$ are circles, corresponding to complex reductive groups on the target of \eqref{eq:MT}. Morphisms are bordisms among circles. In particular, we have those presented in figure \ref{fig:cyl}.

\begin{figure}[ht!]  
\begin{center} 
\includegraphics[scale=0.8]{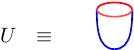}   \ \ \ \ \ \includegraphics[scale=0.8]{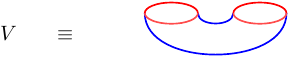}  \ \ \ \ \ \  \ \ \ \ \ $W\ \equiv\ \ \ \ \ \ \ $\includegraphics[scale=0.4]{cypop.pdf} 
\caption{\small These are the bordisms allowed when the target ${\cal C}$ is holomorphic symplectic. When dealing with hyperk$\ddot{\text{a}}$hler quotients, instead, we need to remove the identity bordism, $V$, for the function to be well-defined. This was addressed in a previous work by the same author, \cite{P}.}         
\label{fig:cyl}
\end{center}  
\end{figure}  

In the MT setup, a crucial part is played by the universal centraliser, $Z_{_G}$, defined as follows  

\begin{equation}  
Z_{_G}\ =\ \eta_{_G}\left( S^{^2}\right).  
\label{eq:ZG}   
\end{equation}

The definition of \eqref{eq:ZG} requires that of the Kostant-Wittaker symplectic reduction. Take a subgroup $N\subset G$, such that 

\begin{equation}  
N/[N,N]\ \simeq\ \bigoplus_{_{\{\alpha_{_i}\}}}\mathbb{C}_{_{\alpha_{_i}}}.   
\end{equation}

Then, define a map   

\begin{equation} 
\chi: N\ \longrightarrow\ \mathbb{C}, 
\end{equation}
and    

\begin{equation}  
KW_{_N}\ \overset{def.}{=}\ \mu^{^{-1}}(\chi)/N.             
\end{equation}  

Applying this to the Moore-Tachikawa setup, we get 

\begin{equation}  
KW_{_G} \left(T^*G\right)\ \simeq\ G\times K  
\ \ \ \ \ \ \  , \ \ \ \ \ \ \ \ 
KW_{_{C^{^{\text{V}}}}}(G\times K)\ \simeq\ Z_{_G}  
\label{eq:Langlands}
\end{equation}

\begin{equation}  
KW_{_{G\times G}} \left(T^*G\right)\ \simeq\  Z_{_G}.  
\end{equation}

Most importantly, \eqref{eq:Langlands} is a statement of Langlands duality, where $K$ denotes a Kostant slice. In \eqref{eq:Langlands}, the first equivalence is taken by Moore and Tachikawa as an axiom needed for constructing the 2D TFT in question. However, the most important application of the Moore-Tachikawa varieties consists in the case where the target of the functor \eqref{eq:MT} is a hyperk$\ddot{\text{a}}$hler quotient, which is indeed the case when attempting to describe Higgs and Coulomb branches with this formalism. 

In such case, there is no direct counterpart of the identity element in the source category, Bo$_{_2}$. This matter was already dealt with in a previous work by the same author \cite{P}. 

This brief overview of MT varieties is in the context of the $A_{_1}$-quiver singularity. As explained by Moore and Tachikawa in their original work, the 2D TFT prescription can be suitably extented to the case of $A_{_n}$-singularities by enriching the bordisms by adding punctures carrying extra data. The extra data in question would be a homomorphism

\begin{equation}  
\rho:\ \mathfrak{sl}(2)\ \longrightarrow\ \mathfrak{g}_{_{\mathbb{C}}}.  
\end{equation}  

This variety is characterised by a Hamiltonian action of 

\begin{equation}    
G_{_{\mathbb{C}}}\ \times\ Z(\rho),  
\end{equation}  
where $Z(\rho)$ is the centraliser of $\rho(SL(2))$ inside $G_{_{\mathbb{C}}}$.

Specifically, they would read    

\begin{equation}  
\eta_{_{G_{_{\mathbb{C}}}}}\left(U\right)\ =\ G_{_{\mathbb{C}}}\times S_{_{\rho(e)}}\ \subset\ G_{_{\mathbb{C}}}\times \mathfrak{g}\ \simeq\ T^{^*}G_{_{\mathbb{C}}},   
\end{equation}
where $S_{_{\rho(e)}}$ is a Kostant slice.

\begin{figure}[ht!]  
\begin{center} 
\includegraphics[scale=1.5]{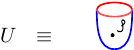}   
\caption{\small Bordism with puncture for the case of a general $A_{_n}$-singularity.}         
\label{fig:cyl}
\end{center}  
\end{figure}

Albeit being very efficient in describing $A_{_n}$-singularities, the formalism described by Moore and Tachikawa requires further generalisations, and, in turn, new mathematical tools. For example, most recently, in \cite{CM}, the Moore-Tachikawa result is recovered as a particular limit of a 1-shifted symplectic construction. However, the main point of our work is that of highlighting how the Moore-Tachikawa setp provides a useful setup for generalising the functorial formulation of HMS.

The starting point for explaining this is an observation originally put forward by C. Teleman, stating that, in presence of a Hamiltonian group action on the symplectic side, the HMS dual is trivial. Importantly, this does not mean that HMS is wrong. But, rather, that it requires a suitable generalisation, comprising this case. In attempting to do so, he developed fascinating works in the context of Coulomb ranches of 3D ${\cal N}=4$ SCFTs, in turn inspired by Braverman, Finkelberg and Nakajima's mathematical formulation, \cite{BFN}. As explained in Section \ref{sec:6.1}, the HMS dual of the Hamiltonian group action can be nontrivially described in terms of the Rozansky-Witten theory. 

\subsection{The role of Coulomb Branches for classifying 2D TFTs}   \label{sec:5.4}

The work of Xie and Yau \cite{XY}, showed the importance of the role played by Coulomb branches of 3D ${\cal N}=4$ SCFTs in classifying SCFTs.  

In more mathematically-oriented papers, but with similar purposes, \cite{CT, CT1}, Teleman addressed their importance in classifying 2D TFTs.

Combining the insights from these and other authors, and applying them to the Moore-Tachikawa setup, identifying the Coulomb branch for describing the singularity in question, enables to achieve a well-defined functor \eqref{eq:MT}.  This is one of the main statements of the present work, that we will turn to explain in more detail in the following section.

\section{Main technique}     \label{sec:5}    

The present section outlines a major technique for determining possible extensions of the Fukaya categories that are of particular relevance for String Theory setups. \footnote{We refer the interested reader to Appendix \ref{sec:6.2} and \ref{sec:6.3} for further techniques that have been explored in the literature. It instructive to note that all three techniques are intertwined with each other. However, we will only be focusing on one of them for the purpose of this article.}This Section is structured as follows:  

\begin{enumerate}   

\item  Section \ref{sec:6.1} outlines the abelianisation technique, one of the main techniques in addressing HMS for singular varieties. Concretely relying on the quantum mechanics of the cohomology of Rozansky-Witten theory on the B-side, we explain how it induces additional Lagrangian submanifolds on the A-side.

\item   Section \ref{sec:vfc} explains the relation between gauging and deformation quantisation to the construction of virtual fundamental classes. In so doing, we explain why it is unsuitable for the nonabelian case, and why we should expect currently-known techniques to be further extended.
\end{enumerate}

We will see how the works of \cite{MS, MT, XY, CT,CT1, BFN} come together, and how this is related to necessary extensions of HMS in its functorial formulation.

\subsection{Abelianisation}    \label{sec:6.1}

\textbf{Cohomology of Rozansky-Witten theory}  

In order to define the HM dual to a Hamiltonian group action, we need to go to the B-side, namely to the Rozansky-Witten (RW) theory. By reducing the B-side to the quantum mechanics of the cohomology of the RW, we can simply restate the question of identifying the Hilbert space on the circle, $C$, at the base of the cylinder in figure \ref{fig:cyl}.

\begin{figure}[ht!]  
\begin{center} 
\includegraphics[scale=0.6]{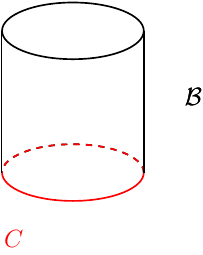}    \ \ \ \ \ \ \ \ \ \  \ \ \ \ \ \  \ \  \ \ \ \ \ \ 
\includegraphics[scale=0.6]{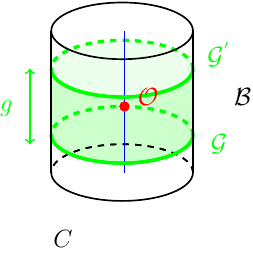} 
\caption{\small The B-side. The QM of the theory living on the circle, $C$. In this case, the moduli space is topologically contractible, and therefore corresponds to the abelian case. The LHS corresponds to the case in which the moduli space is topologically nontrivial, due to the presence of a defect in the bulk of the cylinder. For it to be encoded in a cohomological calculation, we need to consider the moduli space on the union of two copies of $C$, following the prescription outlined in the text.}         
\label{fig:cyl}
\end{center}  
\end{figure}   

The LHS of figure \ref{fig:cyl} denotes the simplest case, when the moduli space is a contractible topological space. Specifically, we can express it as follows  

\begin{equation}  
{\cal M_{_{[C]}}}\ \overset{def.}{=}\ \{(E,X)\},     
\end{equation}   
where $E$ denotes the trivial bundle, and $X$ the zero-section. The resulting Hilbert space therefore reads  

\begin{equation}  
{\cal H}\ =\ H^{^{\bullet}}_{_{G\times U(1)_{_C}}}\left({\cal M_{_{[C]}}}\right).  
\end{equation}  

However, for the case in which an additional defect is placed in the bulk of the cylinder (as denoted on the RHS of figure \ref{fig:cyl}), the space in question is no longer trivially contractible. Such obstruction thereby requires extra care when calculating cohomologies. In particular, we can proceed as follows: take two identical circles (thick, green) and take their distance to be enough to encode the topological defect ${\cal O}$ in between them. Then, the resulting Hilbert space reads  

\begin{equation}  
{\cal H}\ =\ H^{^{\bullet}}_{_{{\cal G}\times {\cal G}^{^{\prime}}\times U(1)_{_C}}}\left({\cal M}_{_{[C\cup C]}}\right)\ \simeq \   H^{^{\bullet}}_{_{{\cal G}\times U(1)_{_C}}}\left({\cal M}_{_{[C\cup C]}}/ {\cal G}^{^{\prime}}\right)\ \hookrightarrow\ H^{^{\bullet}}_{_{T\times U(1)_{_C}}}\left({\cal F}\right)^{^{loc}}\  . 
\label{eq:abelianisation}     
\end{equation}  

The resulting topological space is now topologically non-trivial. The immersion map in \eqref{eq:abelianisation} is the abelianisation process, namely the immersion of the topologically nontrivial cohomology in equivariant localisation. Importantly, the latter gives us the nontrivial B-side that we were looking for, and, at this point we can use the ordinary HMS statement relating cohomologies on the A- and B-sides, namely attempting to identify the corresponding Fukaya category, and, with it, its Lagrangian submanifold objects.

\begin{figure}[ht!]  
\begin{center} 
\includegraphics[scale=0.55]{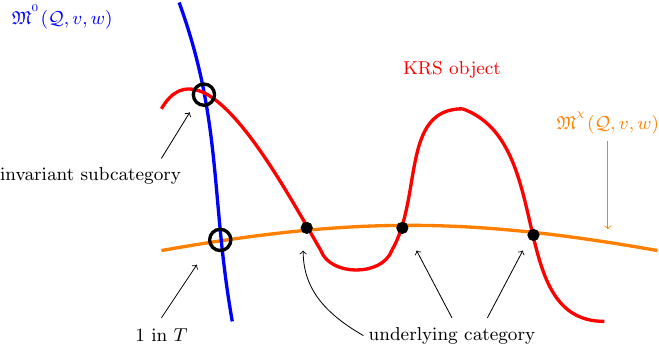}   \ \ \ \   \ \ \ \ \ \ \ \ \ 
\includegraphics[scale=0.55]{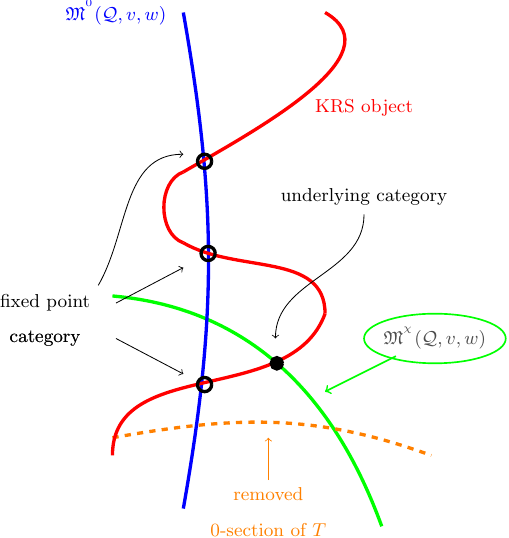} 
\caption{\small The A-side. This picture corresponds to the HMS dual of figure \ref{fig:cyl}, denoting pairwise Lagrangian intersections spanning the BFN space, namely the Coulomb branches of 3D ${\cal N}=4$ SCFTs. The LHS depicts the abelian case, whereas the RHS the non-abelian case. Please refer to the text for a more detailed explanation.}     
\label{fig:cyl}
\end{center}  
\end{figure}

On the A-side, the Lagrangian intersection in the abelian case, describes a coordinate system foliating the BFN space in a preferred way, as shown on the LHS of figure \ref{fig:cyl}. This is because one can replace either Lagrangian with its intersection with a KRS brane, and reconstruct the bulk profile of the brane by Morse flow. This is equivalent to stating that the Lagrangian intersection encodes all necessary information needed on the A-side. From the BFN perspective, this means there is a 1-to-1 correspondence in between the Coulomb branch of the 3D ${\cal N}=4$ SCFT associated to it and the Lagrangian submanifolds constituting the Fukaya category in question, \cite{BFN}.  

On the other hand, in the non-abelian case, we can no longer trade the KRS-brane/Lagrangian intersection for the full objects, precisely because the Lagrangians, and their intersection, as they are are not enough for foliating the whole BFN space, \cite{KRS}. This is pictured on the RHS of figure \ref{fig:cyl}.

Among the main points we wish to highlight is that \eqref{eq:abelianisation} enables to identify a possible preferred foliation of an enlarged B-side, which, by HMS, leads in turn to an enlarged Fukawa category on the A-side. This obviously involves introducing new Lagrangians, whose pairwise-intersection is now able to provide a preferred foliation of the BFN space of the theory in question. 

What has just been reported above is essentially a statement of the power of functoriality applied to HMS in determining the field content of a given quiver gauge theory. Of course, the setups in questions are highly supersymmetric, calling for the need to extend this analysis to fewer supersymmetries to accommodate physics closer to the Standard Model of Particle Physics.

\subsection{Relation to the virtual fundamental class}   \label{sec:vfc}

We now turn to explaining how the abelian/nonabelian cases outlined above are related to the categorification of invariants introduced in Section \ref{sec:2}. In order to do so, we need to set some preliminary notation, mostly taken from \cite{CT2, KRS}. We denote by $\mathfrak{Coh}(\mathcal{X})$ the category of coherent sheaves of modules over a scheme or variety, $\mathcal{X}$.  

For torus action, which is abelian, gauging the Fukaya category of such a variety, $\mathcal{F}(\mathcal{X})$, amounts to enrich it from $\mathfrak{Coh}\left(T_{_{\mathbb{C}}}^{^{\text{V}}}\right)$-module\footnote{Consisting of a 2-category $\mathfrak{Coh}(\mathcal{X}^{^{\text{V}}})$ determined by the map  

\begin{equation}  
\pi:\ \mathcal{X}^{^{\text{V}}}\ \longrightarrow\ T_{_{\mathbb{C}}}^{^{\text{V}}}  
\end{equation}   
whose objects are Lagrangian submanifolds. These are the 2D TFTs at the boundary of the RW-theory (the 3D bulk TFT).} to an object in $\sqrt{\mathfrak{Coh}\ }\left(T^{^*}T_{_{\mathbb{C}}}^{^{\text{V}}}\right)$\footnote{Such objects are the KRS-branes living in the 3D TFT. Such category is the one with $T$-action that sees the germ of the object near the 0-section (drawn in orange in the picture on the left-hand-side of Figure \ref{fig:cyl}.}. In the nonabelian case, though, gauging brings a holomorphic algebraic manifold which is not a cotangent bundle\footnote{The reason for it not being a cotangent bundle as a whole is that it involves two projections.}   

\begin{equation}  
T^{^*}G_{_{\mathbb{C}}}^{^{\text{V}}}\ \equiv\ BFM\left(G^{^{\text{V}}}\right)\ \longleftarrow\ BFM\left(G^{^{\text{V}}}\right)\ \underset{\mathfrak{t}_{_{\mathbb{C}}}/W}{\times}\ \mathfrak{t}_{_{\mathbb{C}}}\ \longrightarrow\ T^{^*}T_{_{\mathbb{C}}}^{^{\text{V}}}  
\label{eq:twoarrows}    
\end{equation}  

The crucial difference between \eqref{eq:twoarrows} and the abelian setup, is that the former does not allow for a direct morphism between  $T^{^*}G_{_{\mathbb{C}}}^{^{\text{V}}}$ and $T^{^*}T_{_{\mathbb{C}}}^{^{\text{V}}}$. As explained in Section \ref{sec:6.2.1}, this means it is not possible to define a SOT by means of any of the techniques outlined in Section \ref{sec:1}. Indeed, the presence of two distinct projections that are not isomorphic to each other in \eqref{eq:twoarrows} implies that, this time, the definition of the KRS-objects involves two distinct  $\sqrt{\mathfrak{Coh}\ }(X)$.

\subsubsection{$\sqrt{\mathfrak{Coh}\ }(X)$}    \label{sec:6.2.1}

A \emph{symmetric complex}   $(\mathbb{E},\theta,\text{or})$ on an Artin stack, $\chi$, is a perfect complex $\mathbb{E}$ of amplitude [-2,0] consisting of the following data:   

\begin{enumerate}  

\item A non-degenerate symmetric form, $\theta$ on $\mathbb{E}$, namely a morphism   

\begin{equation}  
\Theta:\ \theta_{_{\chi}}\ \rightarrow\ \left(\mathbb{E}\otimes\mathbb{E}\right)[-2]  
\end{equation}  
in $D\text{Coh}(X)$, invariant under transport such that the induced morphism  

\begin{equation}  
\mathbb{E}^{^{\text{V}}}\ \rightarrow\ \mathbb{E}[-2]  
\end{equation}    
is an isomorphism.   

\item An orientation, $\text{or}$, of $\textbb{E}$, namely an isomorphism  

\begin{equation}     
\text{or}:\ \mathcal{O}_{_{\mathcal{X}}}\ \xrightarrow{\simeq}\ \text{det}\left(\mathbb{E}\right)  
\end{equation}  
such that   

\begin{equation}  
\text{det}\left(\iota_{_{\theta}}\right)\ =\ \text{or}\circ\ \text{or}^{^{\text{V}}}  
\end{equation}

\end{enumerate}

For any symmetric complex $\mathbb{E}$, there is a quadratic function  
\begin{equation}  
q_{_{\mathbb{E}}}:\ \mathfrak{C}_{_{\mathbb{E}}}\ \rightarrow\ \mathbb{A}_{_{\mathcal{X}}}^{^1}
\label{eq:qe}
\end{equation}  
with $\mathfrak{C}_{_{\mathbb{E}}}$ denoting the virtual normal cone of $\mathbb{E}$. When $\mathbb{E}=E[1]$ for SO$(*,\mathbb{C})$-bundle, $E$, the function $q_{_{\mathbb{E}}}$ is given by the quadratic form on $E$. Denoting by $\mathfrak{Q}(\mathbb{E})$ the 0-locus of the quadratic function \eqref{eq:qe}. The SOT  

\begin{equation}  
\phi:\ \mathbb{E}\ \rightarrow\ L_{_f}  
\label{eq:isomsot}
\end{equation}  
for the Deligne-Mumford morphism

\begin{equation}  
f:\ \mathcal{X}\ \rightarrow\ \mathcal{Y}  
\end{equation}   
induces an embedding  

\begin{equation}  
a:\ \mathfrak{C}_{_f}\ \hookrightarrow\ \mathfrak{Q}(\mathbb{E})  
\end{equation}  
leading to the square root virtual pullback map

\begin{equation}  
\sqrt{f!\ }:\ CH_{_{\bullet}}(\mathcal{Y})\ \rightarrow\ CH_{_{\bullet}}\left(\mathfrak{C}_{_f}\right)\ \rightarrow\ CH_{_{\bullet}}(\mathfrak{Q}(\mathbb{E}))\ \xrightarrow{\sqrt{0^{^!}_{_{\mathfrak{Q}(\mathbb{E})}}\ }}\ CH_{_{\bullet}}(\mathcal{X})  
\end{equation}

When $\mathcal{X}$ is a separated Deligne-Mumford stack over $\mathbb{C}$ with structure morphism  

\begin{equation}   
a_{_{\mathcal{X}}}:\ \mathcal{X}\ \rightarrow\ \text{Spec}\ \mathbb{C}  
\end{equation} 
and an isomorphic SOT, \eqref{eq:isomsot}, the virtual fundamental class is defined as follows

\begin{equation}    
[\mathcal{X}]^{^{\text{vir}}}\ \overset{\text{def.}}{=}\ \sqrt{a^{^!}_{_{\mathcal{X}}}\ }\left[\text{Spec}\ \mathbb{C} \right] \ \in\ CH_{_{\text{Vdim}(\mathcal{X})}}(\mathcal{X})   
\label{eq:virtfundcl}
\end{equation} 

As long as we deal with the abelian setup (cf. the left-hand-side picture in Figure \ref{fig:cyl}), we can always define \eqref{eq:virtfundcl}, and the degeneracy formula can in turn be achieved since the two sides will be related by deformation quantisation. In such case, we can apply the shifted symplectic structure technique to DG-modules and calculate invariants without the need to constrain the analysis to ideal sheaves.   

\subsection{Main points}  

From our treatment, we are therefore led to make the following observations:    

\begin{enumerate}  

\item  We need to understand how the virtual fundamental class can be defined in the nonabelian case, namely where two distinct $\sqrt{\mathfrak{Coh} \ }(X)$ are involved. The main issue being that the two sides are not related by a single deformation-quantisation step.  

\item How to recover the abelianised version of 1. It is reasonable to expect that this should introduce new sheaves.  

\item Physically, this should correspond to new fields in the corresponding String Theory setup.    

\item Shifted symplectic structures do not provide the ultimate answer to address this question, due to the constraining requirements on the sheaves involved for the calculation of invariants. Even upon relaxing the ideal sheaf assumption for achieving the degeneracy formula, we still cannot claim anything regarding the nonabelian setup. So far, only the abelian case can be addressed with such tool.   The A-side of homological mirror symmetry therefore requires further advancements.

\item  Abelianisation is the ultimate goal, but there is clearly a formal gap to be filled, which should be filled, and that should clearly provide very promising useful insights for, both furthering the String Theory realisation of effective field theories, as well as the development of mathematical tools for enumative geometry and homological mirror symmetry.

\end{enumerate}

We are currently working towards identifying the nonabelian counterpart of \eqref{eq:virtfundcl}, understanding its mathematical structure, and the implications of its abelianisation to the degeracy formula as well as to its D-brane realisation. We plan to report any advancement in this regard in due course.

\section{Conclusions and Outlook}

The essential theme in this article is the importance of further generalising the mathematical structure underlying functorial HMS, mostly motivated from its applicability to String Theory, with quiver gauge theories playing a crucial role in the setting. However, as we have seen, generalising this construction also constitutes a promising setting for pure mathematics advancements in their own right.   

Our main finding consists in having related stability and transversality in the degeneracy formula for non-ideal sheaves to nonabelian gauging in the RW-theory setup.

The present work was structured as follows: Section \ref{sec:1} introduced symmetric obstruction theories as shifted symplectic structures, \cite{DO, PTVV}. Section \ref{sec:2} briefly overviewed a recent work by Katzarkov, Kontsevich and Sheshmani, \cite{KKS}, pertaining the categorification of Donaldson-Thomas (DT) invariants for cases involving non-smooth moduli spaces, relative CY structures and spherical functors. The example we outlined falls within the theme of categorification of AG. We highlighted the types of mathematical structures we needed to apply to achieve the degeneracy formula, and how it relates to deformation quantisation. In Section \ref{sec:3} we then turned to the topological string theory formulation of Section \ref{sec:2}, mostly related to \cite{Pan,AOSV}. Here we stressed the role played by different brane types in the braneweb for which the topological partition function is being calculated, and how they are encoded in a 2-functor corresponding to Gromov-Witten (GW) invariants\footnote{DT-invariants are an equivariant formulation of GW-invariants. Calculations coincide for CY3-folds.}. Cobordism calculations, though, highlighted that such invariants restrict to critical points of a topological Morse function. More interesting configurations should be expected to arise when attempting to analyse geometric structures arising at arbitrary points of the flow. In Section \ref{sec:4}, we applied the tools outlined in the previous sections to a specific setup of interest for its applicability to String Theory, namely that of Moore-Tachikawa varieties, \cite{MT,MS,P,CM}. Particular emphasis was placed on the role of Coulomb branches for classifying 2D Topological Field Theories (TFTs), \cite{XY,CT,CT1,BFN}. The reason for doing so is that such 2D TFTs constitute the boundary of a 3D TFT described by Rozansky and Witten for describing the category of branes associated to the B-model and their deformation. Section \ref{sec:5} turned to a major technique used for extending HMS for singular varieties, namely that of abelianisation. In so doing, we explained the relation between gauging and deformation quantisation, \eqref{eq:defquant}, and the difference between the abelian and the nonabelian setup, highlighting how this relates to the definition of the virtual fundamental class in the degeneracy formula. We concluded by outlining some main open questions and currently ongoing work by the same author towards extending the formalism in question.  
We plan to report any advancement in this regard in due course.

\section*{Acknowledgements}

The author wishes to thank Gregory Moore, Cumrun Vafa, and Edward Witten for insightful discussions and questions raised at different stages of the present work. Early stages of it have been presented at the Chern Mathematics Institute (Tianjin), KITS Beijing, Fudan University Maths Department (Shanghai), and NYU in Abu Dhabi in occasion of Strings 2025. An almost ultimated version was presented as a plenary talk at the Geometry and Physics Workshop in January 2025 at TSIMF, Sanya, organised by Professor Shing-Tung Yau and Professor Bong Lian. Its completed version was first presented at a Joint Tsinghua-BIMSA Seminar on Symplectic Geometry in June 2025. The author also acknowledges hospitality from IMSA, Miami, in occasion of the Homological Mirror Symmetry Conference in 2025 during the concluding stage of this work. Last but not least, the author wishes to thank the anonymous referee, for insightful comments and suggestions, that helped towards improving this work towards its present form. The author's research is funded by SIMIS. 

\appendix

\section{Essential Concise Background}  \label{sec:A}

For completeness, this appendix provides some succinct preliminary background on HMS and its functorial formulation, albeit we refer the interested reader to the pioneering works of Kontsevich, \cite{Kont}, and Strominger, Yau and Zaslow, \cite{SYZ}, for detailed references, as well as the extremely rich literature that followed throughout the past few decades. Most of the terminology reported here is often referred to throughout the core part of the present article. This appendix is structured as follows: 

\begin{enumerate}  

\item   Section \ref{sec:HMS} presents the original categorical equivalence formulation of HMS, \cite{Kont}. The original formulation comprised CY three-fold varieties.

\item   Section \ref{sec:FHMS} introduces the functorial formulation, of which the categorical equivalence is a particular limit, \cite{SYZ}.

\item  Section \ref{sec:lgcyHMS} explains one of the subsequent developments of the correspondence, namely the Landau-Ginzburg/Calabi-Yau (LG/CY)-correspondence. The importance of this is due to the fact that the LG-model is known to be dual to Fano varieties. Therefore, the LG/CY-correspondence provides a useful advancement in the formulation of HMS bridging between different types of varieties. When combined with Tyurin degenerations\footnote{As explained in Section \ref{sec:2}.},

\item   Last but not least, Section \ref{sec:MFHMS} reproduces some key features of the matrix factorisation (MF) for the pair of pants, \cite{She}. The reason for adding this specific construction is the importance of the three-punctured sphere in the present treatment. Complementary to this is the importance of this setup for introducing the notion of Lagrangian immersion, discussed in Section \ref{sec:5.3}.

\end{enumerate}

\subsection{Homological Mirror Symmetry (HMS)}  \label{sec:HMS}   

Homological Mirror Symmetry (HMS) is the mathematical formulation of a correspondence conjectured by Witten, \cite{Wit}, motivated by String Theory, relating symplectic geometry and algebraic geometry. Kontsevich's original formulation, \cite{Kont}, can be briefly summarised as follows

\begin{equation}  
D^{^b}\text{Fuk}(X)\ \simeq\ D^{^b}\text{Coh}(\hat X)  
\label{eq:cateq}
\end{equation}
where the LHS denotes the derived bounded Fukaya category\footnote{With objects being Lagrnagian submanifolds equipped with local brane structure, whereas morphisms are Floer cohomologies, \cite{PS}.}of $X$, and the RHS is the derived bounded category of coherent sheaves\footnote{Whose objects are skyscraper sheaves, whereas morphisms are Ext-groups.} of $\hat X$. Given that Calabi-Yau (CY) manifolds are, both,  complex and symplectic, \eqref{eq:cateq} provides a way of relating different CY varieties, thereby leading to inevitable interest with regard to its String Theory applicability.

\subsection{Functorial HMS}     \label{sec:FHMS}

Strominger, Yau and Zaslow \cite{SYZ} promoted \eqref{eq:cateq} to a functorial formulation by means of Fourier-Mukai transforms 

\begin{equation}  
{\cal F}: D^{^b}\text{Fuk}(X)\ \longrightarrow\ D^{^b}\text{Coh}(\hat X).  
\label{eq:cateq}
\end{equation}  

For the case in which $X,\hat X$ are mirror CY manifolds, \eqref{eq:cateq} is essentially stating that given a certain symplectic manifold, we can obtain a dual K$\ddot{\text{a}}$hler manifold, belonging to another CY variety. It is known that, from dimensional reduction of 10 D string theory, there is a rich web of dualities arising according to the specific properties of the manifold on which compactification is performed. 

The functorial formulation provided by \cite{SYZ}, later generalised to other manifolds, opens new interesting perspectives to explore potential generalisations and extensions of the original HMS formulation by Kontsevich \cite{Kont}. Specifically, the idea of introducing Lagnangian immersions, as explained in the construction by Sheridan for the case of the pair of pants, \cite{She}, as we shall see later on in this report.

\subsection{The LG/CY correspondence}  \label{sec:lgcyHMS}

A Landau-Ginzburg (LG) model is defined by the pair $(X,{\cal W})$, with $X$ a noncompact complete K$\ddot{\text{a}}$hler manifold, ${\cal W}$ a holomorphic function on $X$. 

In Witten's GLSM, the LG/CY correspondence arises from a variation of the GIT quotient. 

Take $w_{_1},...,w_{_N}$ coprime integers and $x_{_1},...,x_{_N}$ variables of degree $w_{_1},...,w_{_N}$. Let ${\cal W}(x_{_1},...,x_{_N})$ be a weighted homogeneous polynomial of degree $d$, featuring an isolated critical point only at the origin. 
Assuming the CY condition  

\begin{equation}  
d\ =\ w_{_1}+...+w_{_N}  
\end{equation}  
and the Gorenstein condition, namely that $w_{_j}$ divides $d$ $\forall \ 1\le\ j\ N$.  

Two objects can be defined:  

\begin{itemize}

\item  The LG orbifold $(\mathbb{C}^{^N},{\cal W},\mu_{_d})$, corresponding to the space $\mathbb{C}^{^N}$ equipped with an additional action of  

\begin{equation}  
\begin{aligned}
&\mu_{_d}\ =\ \{g\in\mathbb{C}^{^{\times}}|g^{^d}=1\}\\   
&(x_{_1},...,x_{_N})\ \mapsto\ (g^{^{w_{_1}}}x_{_1},...,g^{^{w_{_N}}}x_{_N}),      
\end{aligned}  
\end{equation}
and a $\mu_{_d}$-invariant potential  

\begin{equation}  
{\cal W}:\ \mathbb{C}^{^N}\ \longrightarrow\ \mathbb{C}.  
\end{equation}

\item  The CY hypersurface  $X_{_{\cal W}}\ =\ \{{\cal W}=0\}$ in the weighted projective space 

\begin{equation}  
\mathbb{P}(\underline{w})\ =\ \mathbb{P}(w_{_1},...,w_{_N}),    
\end{equation}
which is quasi-smooth, namely smooth as a stack.

These two models arise from a modified GIT-quotient

\end{itemize}  

\subsection{MF of the pair of pants}   \label{sec:MFHMS}

\begin{equation}   
{\cal P}^{^n}\ \ \ \overset{\text{HMS}}{<-------->}\ \ \ \left(\text{Spec}(R), {\cal W}\right),   
\end{equation}   
where the RHS is a LG model, consisting of $R\equiv \mathbb{C}[\tilde M]$, and ${\cal W}\equiv z^{^{e[n+2]}}$. There is a natural action of $\mathbb{T}$ on $R$ that preserves ${\cal W}$, specifically   

\begin{equation}   
\mathbb{T}\ \overset{def.}{=}\ \text{Hom}\left(M, \mathbb{C}^{^*}\right).  
\end{equation}  

The B-model on $\left(\text{Spec}(R), {\cal W}\right)$ is given by the triangulated category of singularities of ${\cal W}^{^{-1}}(0)$, the latter being quasi-equivalent to MF$(R,{\cal W})$ 

\begin{equation}   
L^{^n}\ \ \ \overset{\text{HMS}}{<-------->}\ \ \ {\cal O}_{_o}\ \in\ D_{_{sing}}^{^b}\left({\cal W}^{^{-1}}(0)\right). 
\end{equation}

HMS requires the categorical equivalence between the minimal $A_{_{\infty}}$-model generated by the immersed Lagrangian, ${\cal A}$, and 

\begin{equation}  
{\cal B}\ \overset{def.}{=}\ \text{Hom}^{^*}_{_{\text{MF}(R,{\cal W})}}\left({\cal O}_{_o},{\cal O}_{_o}\right).  
\end{equation}   

It is instructive to see how the $\mathbb{T}$-action enters the picture. 

MF corresponding to ${\cal O}_{_o}$ is the Koszul resolution of ${\cal O}_{_o}$  

\begin{equation}  
R\otimes \Lambda^{^*}\tilde M  
\end{equation}    
with the deformed differential 

\begin{equation}  
\delta\ \overset{def.}{=}\ \iota_{_n}\ +\ v\wedge   
\end{equation}  
where 

\begin{equation}  
\begin{aligned}
u\ &\overset{def.}{=}\ \sum_{_{j=1}}^{^{n+2}}z_{_j}\theta_{_j}^{^{\text{V}}}\ \in\ R\otimes\ \Lambda^{^*}\tilde M^{^{\text{V}}}\\
v\ &\overset{def.}{=}\ \sum_{_{j=1}}^{^{n+2}}a_{_j}\frac{{\cal W}}{z_{_j}}\theta_{_j}\ \in\ R\otimes\ \Lambda^{^*}\tilde M.     
\end{aligned}  
\end{equation}     

Putting things together, this leads to the following final expression

\begin{equation}  
\text{MF}\ \equiv\ \left(R<\theta_{_1}, ..., \theta_{_{n+2}}>,\delta\right)  
\end{equation}    
where  

\begin{equation}     
\delta\ =\ \sum_{_{j=1}}^{^{n+2}}z_{_j}\ \left(\frac{\partial}{\theta_{_j}}+a_{_j}\frac{{\cal W}}{z_{_j}}\theta_{_j}\ \in\ R\otimes\ \Lambda^{^*}\tilde M \right)     
\end{equation}

\begin{equation}  
\text{End}\left(\theta_{_o}\right)\ =\ R\otimes\Lambda^{^*}\tilde M^{^{\text{V}}}\otimes \Lambda^{^*}\tilde M.  
\end{equation}  

The latter can be thought of as the commutative algebra of differential operators  

\begin{equation}   
{\cal B}\ \equiv\ R\left<\theta_{_1},...,\theta_{_{n+2}},\frac{\partial}{\partial\theta_{_1}},...,\frac{\partial}{\partial\theta_{_{n+2}}}\right>,
\end{equation}
with the differential 

\begin{equation}   
d\ \overset{def.}{=}\ [\delta, -]  
\end{equation}

\section{Stability of Higgs bundles}                  \label{sec:B}    

A Higgs bundle is a holomorphic bundle $\left(E,\overline\partial_{_E}\right)$ and a holomorphic section $\phi$ of End$(E)\otimes\ {\cal K}_{_X}$, where   

\begin{equation}  
\overline\partial_{_E}:\ \Gamma(X,E)\ \rightarrow\ \Gamma\left(X,E\otimes\Omega^{^{(0,1)}}(x)\right),     
\end{equation}    
and $\Gamma(X,E)$ denotes the space of all sections of $E$.

A sub-bundle $F$ of $E$ is $\phi$-invariant if $\phi(F)\ \subset\ F\ \otimes\ {\cal K}_{_X}$. A Higgs bundle $(E,\phi)$ is said to be semistable if for any $\phi$-invariant subbundle $F$, 

\begin{equation}  
\frac{\text{deg}(F)}{\text{rank}(F)}\ \le\ \frac{\text{deg}(E)}{\text{rank}(E)},     
\label{eq:ineq}   
\end{equation}    
and stable if \eqref{eq:ineq} is a strict inequality.

\section{Koszul duality}   \label{sec:6.2}

Originally motivated by several constructions in Geometry and Mathematical Physics, including cohomology of compactifications of certain moduli spaces and cohomology of foliations, Koszul duality was first put forward by Maxim Kontsevich as a conjectured generalised version of Lie theory which could apply to the aforementioned setups, while at the same time reflecting some version of duality theory.

Koszul duality is a duality between quadratic operads, predicted in Formal non-commutative symplectic geometry by Ginzburg and Kapranov.   

For any algebra ${\cal A}$ over a field $\mathbf{k}$ graded by the nonnegative integers with ${\cal A}_{_o}$ finite-dimensional and semi-simple, we can define a Koszul dual $A^!$ which is a quadratic algebra with the same properties. The Koszul complex yields and equivalence between the derived categories of those algebras iff ${\cal A}$ is Koszul. 

A Theorem states that there is a natural isomorphism of operads  

\begin{equation}  
\text{Hom}({\cal P},{\cal Q})\ \simeq\ {\cal P}^!\circ {\cal Q}.  
\end{equation}

For a graded category, ${\cal C}$, equivalent to an $A$-gmod, then the category   

\begin{equation}   
{\cal C}^{!}\ \simeq\ {\cal A}^{!}\text{-mod}  
\end{equation}  
only depends on ${\cal C}$ up to canonical equivalence. TO construct the category ${\cal O}$, we need to choose auxiliary data, determining the finiteness condition s in turn: specifically, we must choose a flavour, $\phi$ (or, equivalently, a $\mathbb{C}^{\times}$-action on $\mathfrak{M}_{_H}$ with weight 1 on the symplectic form) and a stability parameter, $\xi\in(\mathfrak{g}^*)^{^G}$. The element $\xi$ induces an inner grading on this algebra, which is used, in turn, for defining the category ${\cal O}$. Then, performing the GIT quotient, we get  

\begin{equation}  
\mathfrak{M}_{_{H,\xi}}\ =\ \mu^{-1}(0)//_{_{\xi}}G. 
\end{equation}  

Taking the unique closed orbit in the closure of a semi-stable orbit defines a map  

\begin{equation}  
\mathfrak{M}_{_{H,\xi}}\ \rightarrow\ \mathfrak{M}_{_H}. 
\label{eq:map}
\end{equation}

For the case in which \eqref{eq:map} is a resolution of a singularity, we recover a Nakajima quiver variety, as described in \eqref{eq:Mt}. However, in general, this might not be the case. In particular, it might be that \eqref{eq:map} is not the dominant map, thereby motivating the need for a broader formalism in the first place.  

To the data $(G,V,\phi,\xi)$, there are two associated versions of category ${\cal O}$: 

\begin{enumerate}  

\item   ${\cal O}_{_{\text{Higgs}}}$, namely the generic category associated to the flavour $\phi$

\item  ${\cal O}_{_{\text{Coulomb}}}$, the algebraic category ${\cal O}$ for the quantisation of $\mathfrak{M}$ defined by the flavour $\phi$ with integral weight.     

\end{enumerate}  

A Theorem by B. Webster, \cite{Webster:2016rhh}, claims there to be a functor    

\begin{equation}  
\boxed{\ \ \ {\cal O}^{^{!}}_{_{\text{Coulomb}}}\ \longrightarrow\ {\cal O}_{_{\text{Higgs}}}. \color{white}\bigg]\  }    
\end{equation}   

For the case in which $\mathfrak{M}$ is a Nakajima quiver variety, this functor reduces to an equivalence. 

A Conjecture claims that, if $\mathfrak{M}$ is the Higgs branch of 3D ${\cal N}=4$ SYM, then the Koszul dual, $\mathfrak{M}^{^{!}}$ exists and is the Coulomb branch of the same theory (equivalently, the Higgs branch of its mirror).   

\section{Lagrangian immersion}      \label{sec:6.3}

As explained in the previous paragraphs, abelianisation combined with the functoriality of HMS, implies the need of introducing new Lagrangian submanifolds, thereby leading to a larger Fukaya category on the A-side.      
In this regard, plenty of significant work has been carried out in the last decade. For the purpose of the work addressed in this report, and on which my research is currently focusing, the analysis performed by Nick Sheridan in the context of HMS for the pair of pants, \cite{She}, is particularly insightful.

\begin{figure}[ht!]  
\begin{center} 
\includegraphics[scale=0.8]{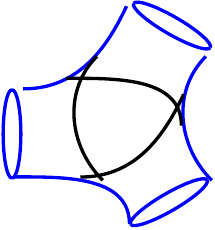}   
\caption{\small In blue, the bordism corresponding to the Higgs branch in the 2D TFT formalism of Moore and Tachikawa, with suitable adaptation to treat Sheridan's construction of the A-side of a HMS dual pair. The black lines denote the embedding of the missing triple self-intersecting Lagrangian to be added to the original Fukaya category associated to the pair of pants.}
\label{fig:POP}  
\end{center}  
\end{figure}

As already mentioned in the previous section, following the Moore-Tachikawa formalism, the bordism associated to the pair of pants, featuring as a homomorphism in Bo$_{_2}$, corresponds to the Higgs branch of a quiver variety with categorical group action.

\begin{equation}
 L^{^n}: S^{^n}\ \rightarrow\ {\cal P}^{^n}\ \ \ \ ,\ \ \ \ \ {\cal P}^{^n}\ =\ \left\{\sum_{_{j=1}}^{^{n+2}}z_{_j}=0\right\}\ \subset\ \mathbb{C}^{^{n+2}} \backslash\ \bigcup_{_j}\ \left\{z_{_j}=0\right\}.  
 \end{equation}

 Define
 \begin{equation}
\tilde X^{^n}\ \overset{def.}{=}\ \left(\left\{z_{_1}^{^{n+2}}+...+z_{_{n+2}}^{^{n+2}}=0\right\}\ \subset\ \mathbb{CP}^{^{n+1}}\right)\ \cap\ \left(\left(\mathbb{C}^{^*}\right)^{^{n+1}}\ \subset\ \mathbb{CP}^{^{n+1}}\right)
\end{equation}   

\begin{equation} 
\text{and its dual}\ \ \ \ \ \ Y^{^n}\ \overset{def.}{=}\ \left(\left\{{\cal W}=0\right\}\ \subset\ \mathbb{CP}^{^{n+1}}\right)\ \ \ , \ \ \ \text{with}\ \ {\cal W}\ \overset{def.}{=}\ z_{_1}z_{_2}...z_{_{n+2}}
\end{equation}   
a singular variety with the natural action of

\begin{equation}
G_{_n}\ \overset{def.}{=}\ \left(\mathbb{Z}_{_{n+2}}\right)^{^{n+2}}/\mathbb{Z}_{_{n+2}}.  
\end{equation}  

A Theorem in \cite{She} states that there exists a fully-faithful $A_{_{\infty}}$-embedding

\begin{equation} 
\boxed{\ \ \ \color{white}\bigg]\color{black}\text{Perf}_{_{G_{_n}}}(Y^{^n})\ \hookrightarrow\ D^{^{\pi}}\text{Fuk}(\tilde X^{^n})\ \ \ \ }         
\end{equation}  
conjectured to be a categorical equivalence.

The deformation space of a Lagrangian immersion is bigger w.r.t. its smoothing, and covers of a local family of Lagrangians. For the case of the pair of pants, Seidel showed that 

\begin{equation}  
CF(L,L)\ =\ \text{Span}\left\{1,X,Y,Z,\overline{X},\overline{Y},\overline{Z},\text{pt}\right\}.  
\end{equation}  

The local moduli is $\left(\mathbb{C}^{^3},{\cal W}\right)$, where ${\cal W}=xyz$. To compute the mirror and recover HMS, the pair of pants can be compactified to $\mathbb{P}^{^1}_{_{a,b,c}}$.

Denoting by ${\cal L}$ the collection of lifts of the Seidel Lagrangian $\overline{\mathbb{L}}$ in a manifold cover, there is an isomorphism 

\begin{equation}  
\begin{aligned}   
\Phi_{_1}:\ CF\left(\overline{\mathbb{L}},\overline{\mathbb{L}}\right)&\ \longrightarrow\ \text{Hom}\left(P_{_{\overline{\mathbb{L}}}}, P_{_{\overline{\mathbb{L}}}}\right)\\
&p\ \mapsto\ \mu_{_2}^{^{b,0}}(-,p)  
\end{aligned}
\end{equation}

If the image of the following open-closed map contains the unit  

\begin{equation}  
{\cal OC}:\ HH_{_{\bullet}}({\cal L})\ \rightarrow\ QH^{^{\bullet}}(M),     
\end{equation}  
then ${\cal L}$ \emph{split-generates}  $Fuk(M)$.

\end{document}